\documentclass[12pt]{iopart}
\usepackage{iopams}
\usepackage{graphicx}
\usepackage{mathtools}
\usepackage{hyperref}
\usepackage{natbib}
\usepackage{xcolor}
\usepackage[normalem]{ulem}

\newcommand{\orcid}[2]{\href{https://orcid.org/#2}{#1}}

\begin{document}

\title[Sensitivity of present and future detectors]{Sensitivity of present and future detectors across the black-hole binary gravitational wave spectrum}
\author{\orcid{A. R. Kaiser}{0000-0002-3654-980X},$^{1,2}$ \orcid{S. T. McWilliams}{0000-0003-2397-8290}$^{1,2}$}
\address{$^1$Department of Physics and Astronomy, West Virginia University, P.O. Box 6315, Morgantown, WV 26506, USA}
\address{$^2$Center for Gravitational Waves and Cosmology, West Virginia University, Chestnut Ridge Research Building, Morgantown, WV 26505, USA}

\begin{abstract}
    Black-holes are known to span at least 9 orders of magnitude in mass: from the stellar-mass objects observed by the Laser Interferometer Gravitational-Wave Observatory Scientific Collaboration and Virgo Collaboration, to supermassive black-holes like the one observed by the Event Horizon Telescope at the heart of M87.
    Regardless of the mass scale, all of these objects are expected to form binaries and eventually emit observable gravitational radiation, with more massive objects emitting at ever lower gravitational-wave frequencies. 
    We present the tool, \href{https://github.com/ark0015/gwent}{\texttt{gwent}}\footnote{Our code, \texttt{gwent}, is available at \href{https://github.com/ark0015/gwent}{https://github.com/ark0015/gwent} and on \href{https://pypi.org/project/gwent/}{\texttt{PyPI}}}, for modeling the sensitivities of current and future generations of gravitational wave detectors across the entire gravitational-wave spectrum of coalescing black-hole binaries.
    We provide methods to generate sensitivity curves for pulsar timing arrays (PTAs) using a novel realistic PTA sensitivity curve generator \citep{Hazboun2019}, space-based interferometers using adaptive models that can represent a wide range of proposed detector designs \citep{2017AmaroSeoane}, and ground-based interferometers using realistic noise models that can reproduce current \citep{Abbott2016}, second, and third generation designs \citep{Hild2011}, as well as novel variations of the essential design parameters.
    To model the signal from black-hole binaries at any mass scale, we use phenomenological waveforms capable of modeling the inspiral, merger, and ringdown for sources with varying mass ratios and spins \citep{Khan2016,Husa2016}.
    Using this adaptable framework, we produce signal-to-noise ratios for the combination of any modeled parameter, associated with either the detector or the source. 
    By allowing variation across each detector and source parameter, we can pinpoint the most important factors to determining the optimal performance for particular instrument designs. 
    The adaptability of our detector and signal models can easily be extended to new detector designs and other models of gravitational wave signals.
\end{abstract}


\submitto{\CQG}

\section{Introduction}
\label{sec:Intro}
The observation of gravitational waves (GWs) from black-hole binaries (BHBs) allows us to probe the Universe on a wide range of scales.
Stellar black-hole binaries up to hundreds of solar masses can be observed from
the ground, whereas more massive objects can only be observed from space.
Massive binaries that are millions of solar masses require space-based detectors, whereas supermassive binaries that are billions of solar masses require detectors larger than our solar system, and are currently being pursued with pulsar timing arrays (PTAs).

In September of 2015, the Advanced Laser Interferometer Gravitational Wave Observatory (LIGO) first detected GWs \citep{Abbott2016}. 
Now with multiple observing runs under LIGO's and Virgo's belt, low mass compact binary mergers are now being studied en masse.
More massive sources have not yet been observed, but we expect they will be in the near future using different types of GW observations. 
Here we focus on three types of GW observatories based on the extension of established technologies: ground-based and space-based interferometers, and PTAs.
Future ground-based interferometric detectors, like Cosmic Explorer \citep{Reitze2019} or the Einstein Telescope (ET) \citep{Hild2011}, will probe a wider parameter space of low mass mergers relative to Advanced LIGO/Virgo.
Proposed space-based detectors, like the Laser Interferometer Space Antenna (LISA) and the Deci-Hertz Interferometer Gravitational wave Observatory (DECIGO), will probe milli-hertz \citep{2017AmaroSeoane} and deci-hertz frequency GWs \citep{Kawamura2006}, respectively, providing a bridge between currently covered gravitational wavebands.
PTAs are currently attempting to detect low-frequency GWs in the nanohertz regime \citep{Hobbs2010,nanoGWB2018}.

Because of the disparate ways in which each detector's sensitivity curve is constructed, it has been difficult to treat them uniformly.
In the past, PTA representations have been often overly simplistic when compared to interferometer detection sensitivities.
With this paper we wish to provide a concise, broadband comparison of probable observations from a representative sample of detectors in each frequency band, and to provide relevant parametrized expressions so that these examples can be easily extended to alternative designs.
Additionally, this uniform treatment we hope to clarify how to accurately and fairly compare the different methods of GW detection and apply their sensitivities to assess each detector's response to the signal from an inspiraling BHB.

In \S\ref{sec:Backsec} we go through the conventions of instrument sensitivity representation, signal parametrization and characterization, and signal-to-noise ratio (SNR) calculation.
In \S\ref{sec:PTAsec}, \S\ref{sec:LISAsec}, and \S\ref{sec:LIGOsec} we apply the methodology in \S\ref{sec:Backsec} to PTAs, space-based GW detectors, and ground-based GW detectors, respectively.
In \S\ref{sec:Discussion} we discuss our results and future work.

\section{Background}
\label{sec:Backsec}
\subsection{Sensitivity Curves}
    When comparing GW detectors, it is useful to have a relatively quick estimate of their sensitivities. 
    Typically this sensitivity attempts to include approximations to real analysis techniques to better represent the true measurement capabilities of the instrument \citep{MooreCole2015}.
    
    With this in mind, we characterize an instrument's sensitivity using its power spectral density (PSD), or the time-averaged power per unit frequency of the noise, normalized by the instrument's response to a GW. 
    To provide an easy visual comparison of different detectors' sensitivities to various signals, one can convert this normalized PSD to an effective characteristic strain amplitude, which gives the source strain that would be equivalent to the level of instrument noise \citep{Thrane2013}.
    
    \subsubsection{Spectral Density}
        It is first useful to define a few quantities that are frequently used in sensitivity estimation. 
        We refer to the PSD of the detector as $P_{n}(f)$, and therefore the amplitude spectral density is $\sqrt{P_{n}(f)}$. 
        The PSD is often defined as a combination of power laws that are set by the different types of noise inherent to the particular detector; for instance, seismic, thermal, and shot noises for (Advanced) LIGO, acceleration and position noises for LISA, and white (or red) noise timing uncertainties for PTAs \citep{MooreCole2015}. 
        Because each PSD estimate depends on the specific design, we define the parameters that determine specific detectors' performances in the following sections: we discuss PTAs in \S\ref{sec:PTAsec}, space-based interferometers in \S\ref{sec:LISAsec}, and ground-based interferometers in  \S\ref{sec:LIGOsec}.
        
    \subsubsection{Instrument Strain Representations}
        To define a more useful representation of instrument sensitivity, we combine the PSD, $P_{n}(f)$, and response function, $\mathcal{R}(f)$, of the detector to define the effective noise PSD, $S_{n}(f)$ as \citep{Wainstein1970}  
        \begin{equation}{\label{eq:ENSD}}
            S_{n}(f) = \frac{P_{n}(f)}{\mathcal{R}(f)} .
        \end{equation}
        This quantity is sometimes displayed for sensitivity plots, but a more useful quantity is the noise's effective strain amplitude,
        \begin{equation}{\label{eq:instcharstrain}}
            h_{n}(f) = \sqrt{fS_{n}(f)} ,
        \end{equation}
        i.e., the PSD is converted from a spectral density to a strain power, and then square-rooted to yield a strain amplitude. As we will see later, one can visually estimate the SNR by simply comparing the height of the effective strain noise amplitude, $h_{n}(f)$, and the characteristic strain of the source, $h_{c}(f)$, which we define in \S\ref{subsec:SigParam}. 

    \subsubsection{Response Functions}{\label{susubsec:Response}}
        The response function, $\mathcal{R}(f)$, characterizes the frequency dependence of the stretching and squeezing experienced by a detector in the presence of a passing GW.
        Since the detector response also depends on the sky location and polarization of the source, as characterized by the antenna pattern, we are often interested in including this dependence in an averaged sense within $\mathcal{R}(f)$.
        The instrument can additionally have some correlation between the detector arms (or pulsar pairs in the case of PTAs) \citep{Thrane2013}. We will summarize the construction of $\mathcal{R}(f)$ in this subsection, but see \cite{Maggiore2000} for further details on response functions, antenna patterns, and inter-instrument correlations.
        
        For a GW propagating in the $\hat{\Omega}$ direction with transverse orthonormal vectors $\hat{m},\hat{n}$ defined by,
        \begin{align}
            \hat{\Omega} &=(\sin \theta \cos \phi, \sin \theta \sin \phi, \cos \theta) \\ 
            \hat{m} &=(\sin \phi,-\cos \phi, 0) \\ 
            \hat{n} &=(\cos \theta \cos \phi, \cos \theta \sin \phi,-\sin \theta) ,
        \end{align}
        the strain is given by the tensor,
        \begin{align}
            \label{eq:h_t}
            \mathbf{h}(t, \mathbf{x})=h_{+}\left(2\pi f( t- \widehat{\Omega} \cdot \mathbf{x});\iota\right) \mathbf{\epsilon}^{+}(\widehat{\Omega}, \psi)+h_{ \times}\left(2\pi f( t-\widehat{\Omega} \cdot \mathbf{x});\iota\right) \mathbf{\epsilon}^{ \times}(\widehat{\Omega}, \psi) ,
        \end{align}
        made up of the two GW polarizations $+$ and $\times$, where $\mathbf{x}$ is a vector defining the spatial position of the source, $\psi$ is the polarization angle describing a rotation of $\hat{m}$ and $\hat{n}$ about $\widehat{\Omega}$ (i.e., $\hat{m},\hat{n} \rightarrow \hat{u},\hat{v}$) such that $\cos{\psi} = \hat{m}\cdot\hat{u}$, and $\iota$ is the inclination between the propagation direction $\widehat{\Omega}$ and the orbital angular momentum unit vector $\hat{L}$, i.e., $\cos{\iota}=\widehat{\Omega}\cdot\hat{L}$. See figures 6 and 7 of \cite{Hazboun2019} for diagrams of the coordinate system.
        The polarization tensors
        \begin{align}
            \label{eq:PolarizationTensors}
            &\mathbf{\epsilon}^{+}(\widehat{\Omega}, \psi)=\mathbf{e}^{+}(\widehat{\Omega}) \cos 2 \psi-\mathbf{e}^{ \times}(\widehat{\Omega}) \sin 2 \psi \\ 
            &\mathbf{\epsilon}^{ \times}(\widehat{\Omega}, \psi)=\mathbf{e}^{+}(\widehat{\Omega}) \sin 2 \psi+\mathbf{e}^{ \times}(\widehat{\Omega}) \cos 2 \psi ,
        \end{align}
        in equation \ref{eq:h_t} are made up of the tensor products of the transverse basis vectors:
        \begin{align} 
            \label{eq:BasisTensors}
            \mathbf{e}^{+}(\widehat{\Omega}) &=\hat{m} \otimes \hat{m}-\hat{n} \otimes \hat{n} \\ \mathbf{e}^{ \times}(\widehat{\Omega}) &=\hat{m} \otimes \hat{n}+\hat{n} \otimes \hat{m}  
        \end{align}
        \citep{Anholm2009}.
        Equivalently, equation \ref{eq:h_t} can be written in terms of the Fourier transform of $\mathbf{h}(t)$,
        \begin{equation}
            \mathbf{h}(t, \mathbf{x})=\sum_{A=+,\times} \int_{-\infty}^{\infty} \mathrm{d} f e^{i 2 \pi f(t-\widehat{\Omega}\cdot\mathbf{x})} \tilde{h}_{A}(f; \widehat{\Omega},\iota) \mathbf{\epsilon}^{A}(\widehat{\Omega},\psi)
        \end{equation}
        where $\tilde{h}_{+,\times}(f;\iota)$ are the Fourier transforms of $h_{+,\times}(t;\iota)$.
        Therefore the response in the GW detector is
        \begin{align}
            h(t ; \widehat{\Omega}, \iota, \psi)=\int_{-\infty}^{\infty} \mathrm{d} f \tilde{h}(f ; \widehat{\Omega}, \iota, \psi) e^{i 2 \pi f t}
        \end{align}
        where
        \begin{align}{\label{eq:htildef}}
            \tilde{h}(f ; \widehat{\Omega}, \iota, \psi)=F^{+}(f,\widehat{\Omega},\psi) \tilde{h}_{+}(f;\iota)+F^{ \times}(f,\widehat{\Omega},\psi) \tilde{h}_{ \times}(f;\iota) ,
        \end{align}
        is the source strain made up of the two GW polarizations,
        \begin{align}
            &\tilde{h}_{+}(f;\iota) = A(f)\frac{(1+\cos^{2}{\iota})}{2}~e^{i\Psi_{+}(f)} \\
            &\tilde{h}_{\times}(f;\iota) = iA(f)\cos{\iota}~e^{i\Psi_{\times}(f)} ,
        \end{align}
         where $A(f)$ is the strain amplitude and $\Psi_{\times,+}(f)$ is the strain phase \citep{Maggiore2007}.
        $F^{+,\times}(\widehat{\Omega},\psi,f)$ in equation \ref{eq:htildef} is the complete response function, incorporating the frequency, antenna pattern, and polarization dependencies for a specific instrument. We discuss PTA detector response functions in \S\ref{subsubsection:PTAResponse} and interferometer response functions in \S\ref{subsec:TransferFunc}.
        
        Since we are interested here in the typical sensitivity to a source, we generally average over the source sky location, inclination, and polarization,
        \begin{align}
            \left\langle\tilde{h}(f ; \widehat{\Omega}, \iota, \psi) \tilde{h}^{*}(f ; \widehat{\Omega}, \iota, \psi)\right\rangle & = \left\langle F^{+}(\widehat{\Omega},\psi,f)F^{+*}(\widehat{\Omega},\psi,f)\right\rangle
            \left\langle\tilde{h}_{+}(f;\iota)\tilde{h}_{+}^{*}(f;\iota)\right\rangle \\
            &\hspace{5mm}+\left\langle F^{ \times}(\widehat{\Omega},\psi,f) F^{ \times *}(\widehat{\Omega},\psi,f)\right\rangle\left\langle\tilde{h}_{ \times}(f;\iota)\tilde{h}_{ \times}^{*}(f;\iota)\right\rangle \nonumber\\
            & = \left\langle F^{+}(\widehat{\Omega},\psi,f)F^{+*}(\widehat{\Omega},\psi,f)\right\rangle \nonumber\\ &\hspace{5mm}\times\left(\left\langle\tilde{h}_{+}(f;\iota)\tilde{h}_{+}^{*}(f;\iota)\right\rangle + \left\langle\tilde{h}_{ \times}(f;\iota)\tilde{h}_{ \times}^{*}(f;\iota)\right\rangle\right) \nonumber,
        \end{align}
        since $\left\langle F^{+}(f) F^{+*}(f)\right\rangle=\left\langle F^{ \times}(f) F^{\times *}(f)\right\rangle$ \citep{Maggiore2007}.
        Thus,
        \begin{align}
            \fl
            \label{eq:angavgstrain}
            \left\langle\tilde{h}(f ; \widehat{\Omega}, \iota, \psi) \tilde{h}^{*}(f ; \widehat{\Omega}, \iota, \psi)\right\rangle &= \frac{1}{4\pi}\int_{0}^{2 \pi} \mathrm{d}\phi \int_{0}^{\pi} \mathrm{sin}\theta~\mathrm{d}\theta \\
            &\hspace{5mm} \times \int_{0}^{2 \pi} \frac{\mathrm{d} \psi}{2 \pi}F^{+}(\widehat{\Omega},\psi,f)F^{+*}(\widehat{\Omega},\psi,f) \nonumber\\
            &\hspace{5mm} \times\int_{-1}^{1}\frac{\mathrm{d}(\cos{\iota})}{2}
            \left[\left|\tilde{h}_{+}(f;\iota)\right|^{2} + \left|\tilde{h}_{\times}(f;\iota)\right|^{2}\right] \nonumber\\
            & = \frac{4}{5}\,\mathcal{R}(f)\left|\tilde{h}(f)\right|^{2} \nonumber
        \end{align}
        where the final equality defines $\mathcal{R}(f)$, $|\tilde{h}(f)|^{2} = |\tilde{h}_{+}(f)|^{2} + |\tilde{h}_{\times}(f)|^{2}$ is the Fourier strain amplitude, and the factor of $\frac{4}{5}$ comes from angle averaging of the inclination angle
        (see e.g., \cite{Thorne1989}).
        We discuss instrument-specific response functions in the sections that follow, and refer back to this section for general statements.
        
\begin{figure}[!htbp]
            \includegraphics[width=\textwidth]{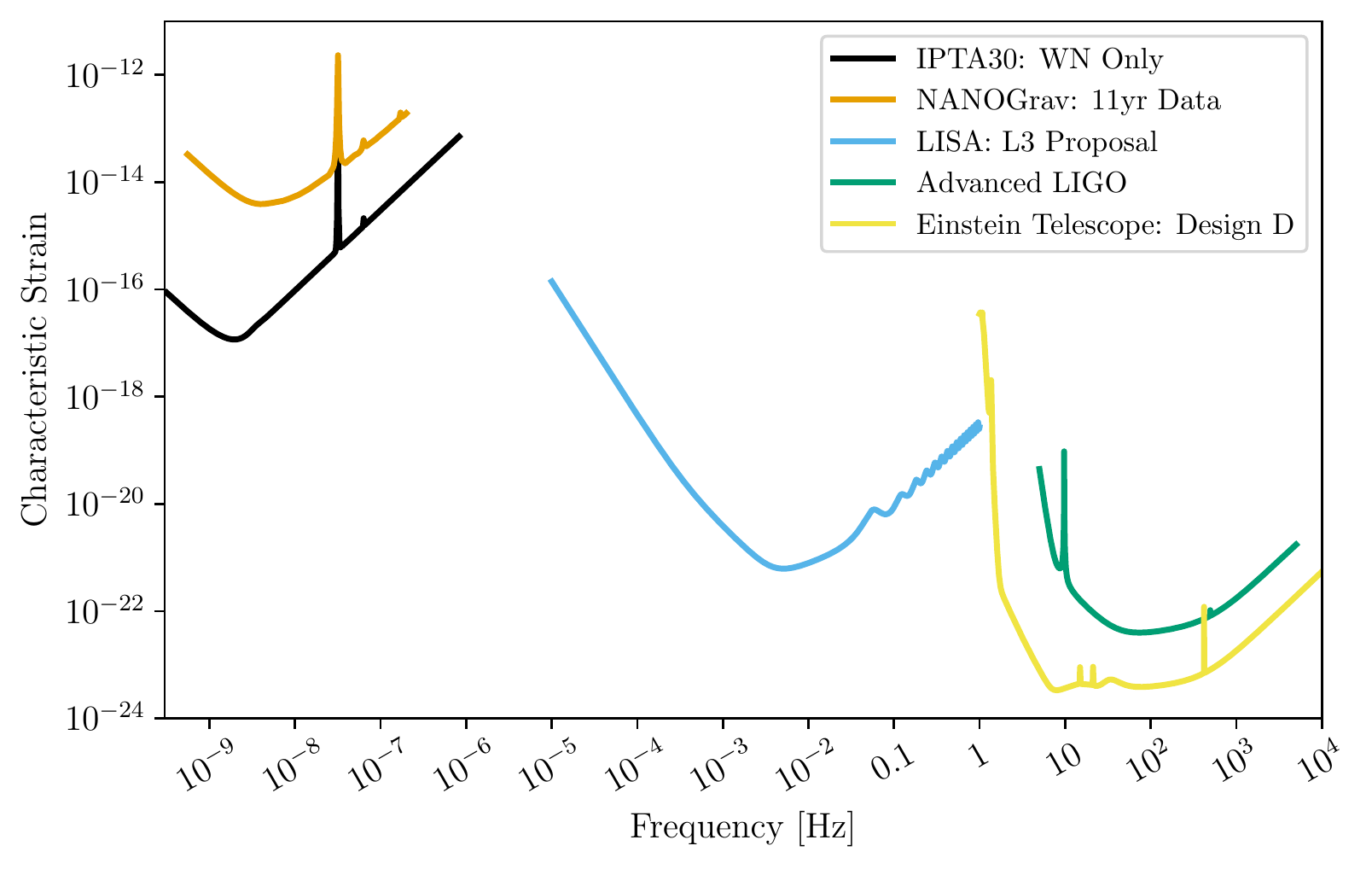}
            \caption{Broadband characteristic strain sensitivity to individual coalescing BHB sources.
            The IPTA30 model uses our white noise only model with $\sigma_{\mathrm{RMS}}=100\mathrm{ns}$, $T_{\mathrm{obs}}=20\mathrm{yrs}$, $N_{\mathrm{p}}=200$, and a one week cadence.
            The NANOGrav curve uses the 11yr data modeled with \texttt{hasasia}.
            See table \ref{tab:PTAparams} and section \S\ref{subsec:sampledPTAs} for more information about each PTA model. 
            The LISA design uses the proposed design from third large mission (L3) of the European Space Agency's Cosmic Vision Plan \citep{2017AmaroSeoane}.
            See section \S\ref{sec:LISAsec} for more specific details on our LISA model.
            The aLIGO and ET curves that represent the design sensitivities are taken from the references in section \S\ref{sec:LIGOsec}. }
            \label{fig:charstrainband}
        \end{figure}

        In figure \ref{fig:charstrainband}, we display the characteristic strain sensitivity curves for the frequency bands of each detector.
        The plotted models are representative of current GW detector designs in the literature and all are generated with \texttt{gwent}.
        We discuss the precise make-up of each sensitivity curve in the relevant sections.

\subsection{Signal Parametrization}{\label{subsec:SigParam}}
    The entire panoply of GW signals includes a variety of source types.
    We choose to focus on the main class of source that is common to all three detector bands: inspiraling BHBs.
    
    \subsubsection{Inspiraling Black-Hole Binaries}
        BHBs are an important source of GW signals detectable with current and future observatories. 
        The loss of energy and angular momentum due to gravitational-wave emission causes the BHB's orbit to decay, eventually driving it to merge \citep{Thorne1989}.
        The coalescence of two black-holes is generally divided into three epochs: inspiral, merger, and ringdown \citep{Flanagan1998}.
        
        The inspiral portion of BHB evolution is the longest of the three stages.
        During the inspiral, the binary emits gravitational radiation and evolves at an initially slow, well known rate that can be accurately modeled using the Post-Newtonian (PN) approximation \citep{Einstein1938}. 
        The merger portion of coalescence is the final collision of the BHB to form a single black-hole.
        During this epoch, the separations are small, and the black-holes are moving at relativistic speeds; numerical relativity simulations are needed to extract information from the binary \citep{SXS2019}.

        Post-merger, the resulting Kerr black-hole continues to emit gravitational radiation through oscillations of its quasinormal modes, an epoch known as the ringdown \citep{Teukolsky1974}.
        During this final phase, the black-hole emits GWs as a superposition of exponentially damped sinusoids.
        
        We use a combination of methods to estimate the signal strength.
        For binaries that do not evolve over the instrument's observation time, we simply utilize the leading order quadrupole contribution in the Fourier domain via the stationary phase approximation \citep{Droz1999}.
        In the case of evolving, or ``chirping" binaries that inspiral (i.e., whose frequency evolves by more than the frequency resolution of the observation) over the detector's observation period, we use a phenomenological model developed in \cite{Husa2016} and \cite{Khan2016}.

        \paragraph{Chirping Sources}
            As described above, coalescing binaries move through several stages of GW emission. 
            In order to combine the PN approximation of the inspiral with the numerical relativity (NR) results for the merger and ringdown, \cite{Husa2016} developed a fairly simple, non-precessing model of the GW signal in the frequency domain.
            Their method uses a hybrid model of inspiral approximates matched with NR simulations to provide an easily adopted, accurate waveform approximate.
            We implemented their method for the first time in a Python package as opposed to using the LALSuite implementation in C \citep{lalsuite}.
            
            The output from IMRPhenomD gives the amplitude and phase of the dominant quadrupolar modes of the GW signal,
            \begin{align}
                \tilde{h}(f ; \Xi, \theta, \phi) &=\tilde{h}_{+}(f ; \Xi, \theta, \phi)-i \tilde{h}_{ \times}(f ; \Xi, \theta, \phi) \\ &=\sum_{m=-2,2} \tilde{h}_{2 m}(f ; \Xi)^{-2} Y_{2 m}(\theta, \phi) ,
            \end{align}
            where $\Xi \in\left(M, \eta, \chi_{1}, \chi_{2}\right)$ are the total mass, symmetric mass ratio, and dimensionless spin parameters, respectively, $(\theta,\phi)$ are the observer's orientation with respect to the orbital angular momentum of the binary, and $Y_{lm}$ are spin-weighted spherical harmonics.
            Since $\tilde{h}_{2(-2)}(f)=\tilde{h}_{22}^{*}(-f)$, we can parametrize the quadrupolar amplitude as 
    	    \begin{align}
                    \tilde{h}_{22}(f ; \Xi)=A(f ; \Xi)~e^{-i \Psi(f ; \Xi)} ,
            \end{align}
            where $A(f ; \Xi)$ is now the geometrized strain amplitude and $\Psi(f ; \Xi)$ is the geometrized strain phase.
    
\begin{figure}[!htbp]
                \includegraphics[width = \textwidth]{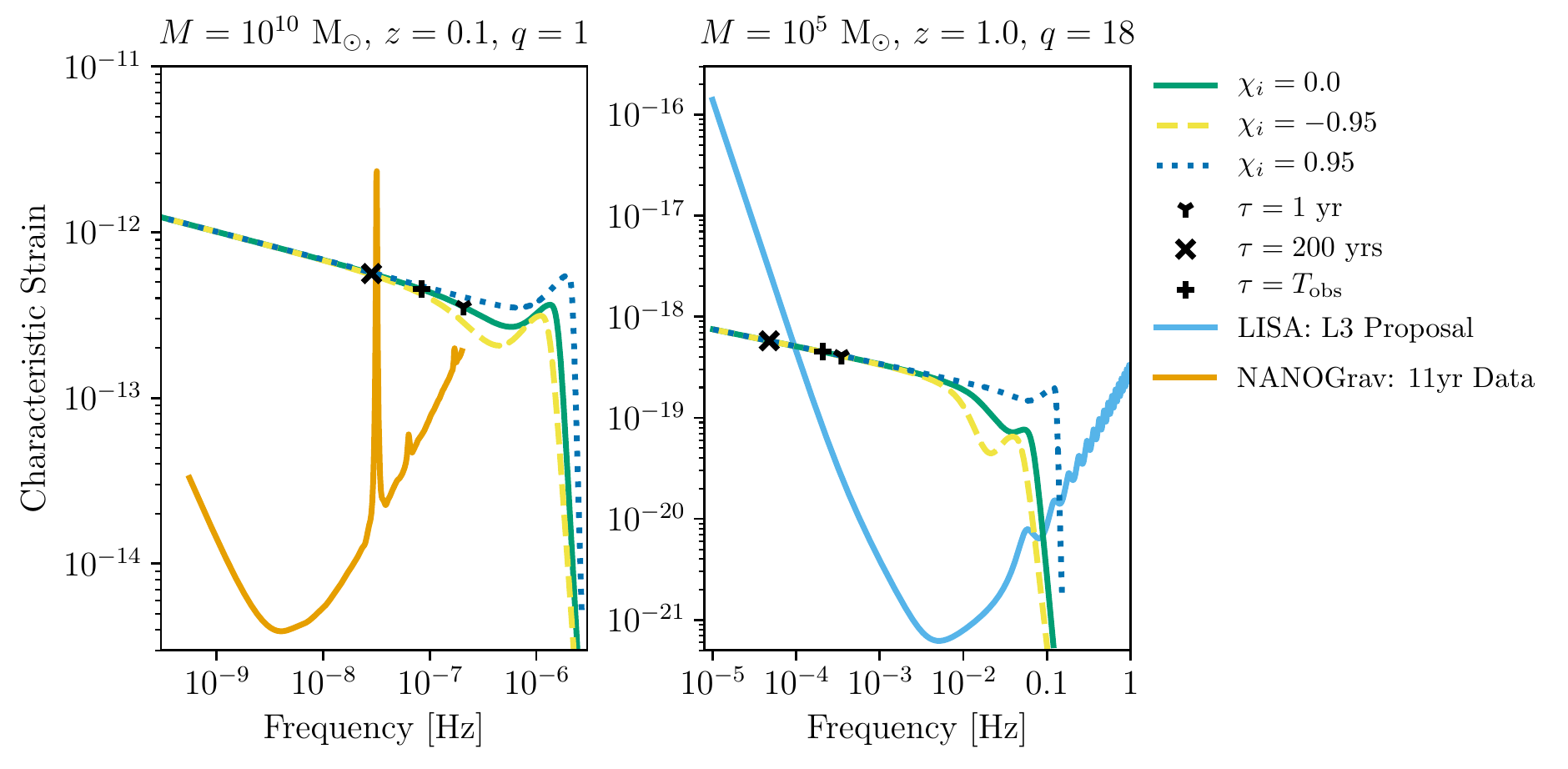}
                \caption{Example GW sources for a PTA source (left) and a LISA source (right) plotted with GW detector sensitivity curves for the NANOGrav 11yr data and the LISA L3 proposal, respectively.
                The definitions for the instrument characteristic strain is found in equation \ref{eq:instcharstrain} and for the source characteristic strain in equation \ref{eq:sourcecharstrain}.
                Each chirping GW source, have the same mass ($M=10^{10} (10^{5}) \mathrm{M}_{\odot}$), redshift ($z=0.1 (1.0)$), and mass ratio ($q=1 (18)$) with three different spin states, no spin ($\chi_{i}=0$), aligned spin ($\chi_{i}=0.95$), and anti-aligned spin ($\chi_{i}=-0.95$). 
                The frequency evolution of the chirping sources, we mark three times to merger (in equation \ref{eq:freqmono}, $\tau = 1\, \mathrm{yr},~T_{\mathrm{obs}}\,\mathrm{yrs},~ \mathrm{and}~200\,\mathrm{yrs}$) for each source.}
                \label{fig:sourceevol}
            \end{figure}

            The IMRPhenomD signal amplitude has the leading order PN contribution factored out through normalizing $A(f ; \Xi)$ by
            \begin{equation}
                \label{eq:leadingPNamp}
                A_{0} = \sqrt{\frac{2\eta}{3\pi^{1/3}}}f^{-7/6} , 
            \end{equation}
            where $\eta= q/(1+q)^{2}$ is the symmetric mass ratio and $q=m_{2}/m_{1}$ is the mass ratio (in our convention $m_{1}\leq m_{2}$) and $m_{1}$ and $m_{2}$ are the individual black-hole masses \citep{Cutler1994}.
            Thus we define the output from IMRPhenomD as $\hat{A}(f ; \Xi)=A(f ; \Xi)/A_{0}$.
            To get the non-normalized amplitude in non-geometrized units and extract the raw Fourier strain amplitude ($|\tilde{h}(f)|$), we assume that as $f \to 0$ (i.e., time until merger $\to \infty$), $|\tilde{h}(f)|$ goes to the leading order post-Newtonian amplitude,
            \begin{align}
                \label{eq:PNamp}
                |\tilde{h}(f)| &= \sqrt{\frac{5}{16\pi}} \frac{M^{2}}{D_{L}(z)}A_{0}(Mf) \hat{A}(Mf ; \Xi) \nonumber \\
                &= \sqrt{\frac{5}{24}} \frac{\mathcal{M}^{5/6}}{\pi^{2/3}D_{L}(z)}f^{-7/6} \hat{A}(f ; \Xi) ,
            \end{align}
            where $D_{\rm L}(z)$ is the source's luminosity distance at a redshift $z$, $M=m_{1}+m_{2}$ is the total mass of the system, and $\mathcal{M} = \eta^{3/5}M$
            is the chirp mass \citep{Cutler1994}. 
            To convert from natural units (i.e., $G=c=1$), one would multiply each $M$ by $G M_{\odot}/c^{3}$ and divide the luminosity distance by the speed of light ($D_{\rm L} = D_{\rm L}/c$). 
            To get the mass in the detector frame, one would redshift the mass to: $M_{z} = (1+z)M$.
            See figure \ref{fig:sourceevol} for examples of chirping source strain given by the IMRPhenomD formalism.
            
        \paragraph{Monochromatic Sources}
            \label{par:monosource}
            For a non-evolving binary, a.k.a. one that remains in the same GW frequency bin over many orbital revolutions, the source strain can be accurately approximated using the quadrupole and stationary-phase approximations \citep{Droz1999}.
            
            To determine whether a source evolves during an observation, either an initial frequency or a time before merger at which the source is observed must be chosen. 
            We select the most sensitive frequency of the detector, $f_{\rm opt}$.
            Thus, the change in GW frequency is,
            \begin{equation}\label{eq:fdot}
                \Dot{f} = \frac{96}{5\pi \mathcal{M}^{2}}(\pi \mathcal{M}f)^{11/3} ,
            \end{equation}
            which is integrated to find the instantaneous GW frequency,
            \begin{equation}{\label{eq:freqmono}}
                f(\tau) = \frac{1}{8\pi \mathcal{M}}\left(\frac{5\mathcal{M}}{\tau}\right)^{3/8} ,
            \end{equation}
            where $\tau=t_{c}-t$ and $t_{c}$ is an integration constant referred to as the time to coalescence.
            By inverting equation \ref{eq:freqmono}, we arrive at the time to merger,
            \begin{equation}
                \tau = \frac{5\mathcal{M}}{256}\big(\pi \mathcal{M}f\big)^{-8/3} .
            \end{equation}
            The change in frequency from the initial observed time until some observation time of the detector, using equation \ref{eq:freqmono} and Taylor expanding, is
            \begin{align}
                \label{eq:deltafreq}
                \Delta f &= f(\tau) - f(\tau-T_{\mathrm{obs}}) \nonumber\\
                &\simeq \frac{3}{64\pi \mathcal{M}}\left(\frac{5\mathcal{M}}{\tau}\right)^{3/8}\frac{T_{\rm obs}}{\tau},
            \end{align}
            where $T_{\rm obs}$ is the detector's observation time.
            Thus, if the frequency has evolved by less than the frequency resolution of the detector ($1/T_{\rm obs}$), i.e., 
            \begin{equation}{\label{eq:freqevol}}
                \Delta f < \frac{1}{T_{\rm obs}}
            \end{equation}
            the source is considered monochromatic.
            
            If we assume, without loss of generality, that $t_{c}=0$ and $t(f)$ is some time until coalescence, and use the assumed optimal frequency $f_{\rm opt}$, we can substitute $t(f_{\rm opt}) = t_{\rm init} = \tau$ into equation \ref{eq:deltafreq}, which gives us the desired time until coalescence for a binary initially observed at the detector's optimal frequency.
            See figure \ref{fig:sourceevol} for an example of how the source strain and frequency evolve with time.
            
            The strain from a monochromatic source is given by,
            \begin{align}{\label{eq:monotimestrain}}
                &h_{+}\left(t ; \iota, \alpha_{0}\right)=h_{0}\left(\frac{1+\cos ^{2} \iota}{2}\right) \cos \left(2 \pi f_{0} t+\alpha_{0}\right)\\
                &h_{ \times}\left(t ; \iota, \alpha_{0}\right)=h_{0} \cos \iota \sin \left(2 \pi f_{0} t+\alpha_{0}\right) ,
            \end{align}{}
            where $h_{0}$ is the approximately constant strain amplitude, $f_{0}$ is the GW frequency of the binary, $\iota$ is the binary's inclination, and the initial phase is $\alpha_{0}$.

            The Fourier transform of equation $h_{+,\times}\left(t ; \iota, \alpha_{0}\right)$
            \begin{align}
                &\tilde{h}_{+}\left(f ; \iota, \alpha_{0}\right)=\frac{h_{0}}{2}\left(\frac{1+\cos ^{2} \iota}{2}\right) \left[e^{i \alpha_{0}} \delta\left(f-f_{0}\right)+e^{-i \alpha_{0}} \delta\left(f+f_{0}\right)\right] \\
                &\tilde{h}_{ \times}\left(f ; \iota, \alpha_{0}\right)= \frac{h_{0}}{2 i} \cos \iota ~ \left[e^{i \alpha_{0}} \delta\left(f-f_{0}\right)-e^{-i \alpha_{0}} \delta\left(f+f_{0}\right)\right] ,
            \end{align}
            can be used in equation \ref{eq:angavgstrain} to get the angle averaged detector response.
        
            If we use the methods in \cite{MooreCole2015} for an inspiralling source in the stationary-phase approximation, and separate out the inclination terms from equation \ref{eq:monotimestrain}, the Fourier strain amplitude is
            \begin{equation}
                \label{eq:fourierstrainmono}
                |\tilde{h}(f)| \simeq \frac{h_{0}}{\sqrt{2\dot{f}}} .
            \end{equation}
            
            This relation allows us to use the amplitude of the very early inspiral portion of the phenomenological model as it asymptotes to the stationary-phase approximation to extract the monochromatic strain amplitude $h_{0}$. 
            As there is little to no numerical difference between this approach and the typical leading-order assumptions made for monochromatic sources, which we lay out below, we choose to use the usual approach because of its computationally simpler and faster properties.

\begin{figure}[!htbp]
                \includegraphics[width = \textwidth]{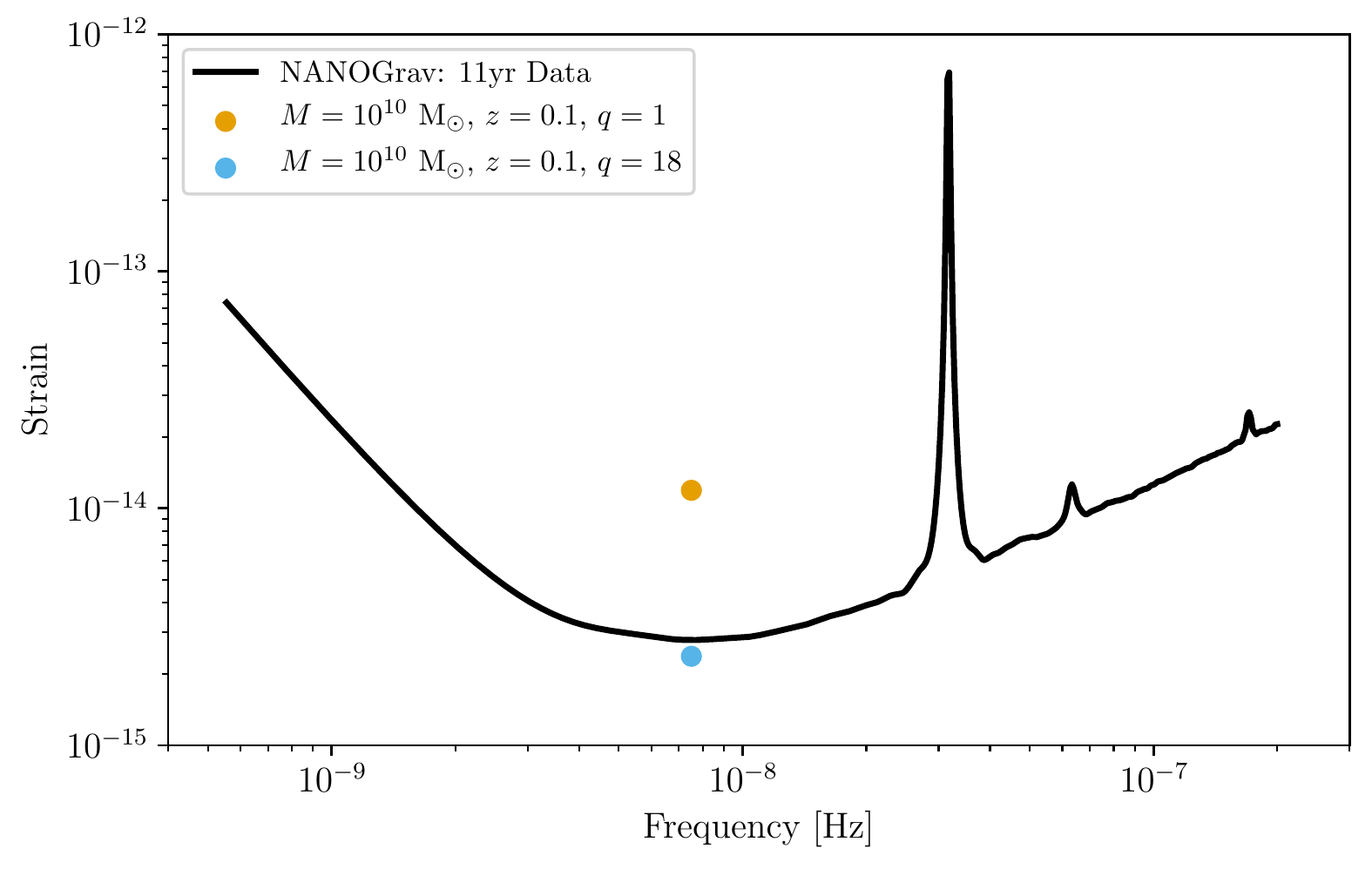}
                \caption{Example monochromatic GW sources for a PTA source plotted with GW detector sensitivity curve for the NANOGrav 11yr data corresponding to a source strain ($h_{0}$) with SNR of one (see eq. \ref{eq:monoSNR}); note that this is not characteristic strain.
                Each source, have the same mass ($M=10^{10} \mathrm{M}_{\odot}$), redshift ($z=0.1$), and different mass ratios ($q=1$ and $q=18$).
                The sources are assumed to be emitting at a GW frequency corresponding to the instrument's most sensitive frequency. 
                }
                \label{fig:monosource}
            \end{figure}

            Thus we use the constant strain amplitude $h_{0}$, given by
            \begin{equation}{\label{eq:monostrain}}
                h_{0} = 4\,\frac{\mathcal{M}^{5/3}}{D_{\rm L}}(\pi f_{0})^{2/3} .
            \end{equation}
            Note that we use the convention of a factor of 4 in the strain amplitude, because of our definitions of equation \ref{eq:monotimestrain}.
            In the literature this factor is sometimes represented by a factor of 2, with a corresponding change in the inclination terms that come out to the same result (see e.g.,  \cite{Thorne1989}).
            If, for our selected optimal frequency, $f_{\rm opt}$, a binary obeys the inequality in equation \ref{eq:freqevol}, then we assume $f_{0}=f_{\rm opt}$ and that the source emits at $f_{\rm opt}$ for the entire observation.
            See figure \ref{fig:monosource} for examples of monochromatic strain amplitudes, $h_{0}$, in the PTA frequency band.
            In the same vein as equation \ref{eq:PNamp}, one must convert the luminosity distance and the chirp mass from natural units.

    \subsubsection{Stochastic Backgrounds}
        Stochastic backgrounds from various GW sources are present in every frequency band \citep{Thrane2013}.
        As a consequence of this, there are many studies on detectability and characterization of backgrounds \citep{Jenet2006,nanoGWB2018,Cornish2017,Parida2016}.
        Because this paper focuses on resolvable, inspiraling BHBs, we treat stochastic backgrounds as an additional noise source. 
        We discuss any specific contributions from stochastic foregrounds in the respective detector sections.

    \subsubsection{Characteristic Strain}
        In order to represent general sources, we choose to absorb the $\psi, \theta$ and $\phi$ terms by using a polarization and orientation averaging (see equation \ref{eq:htildef} and \ref{eq:angavgstrain}).
        This allows us to directly transition between the monochromatic and inspiraling strains.
        In order to get a more useful representation for comparing the signal strain to the sensitivity curves and for calculating the SNR, we use the characteristic strain of the source,
        \begin{equation}{\label{eq:sourcecharstrain}}
            h_{\mathrm{c}}^{2}(f)=4 f^{2}|\tilde{h}(f)|^{2} .
        \end{equation}
        As previously mentioned, one can visually estimate the SNR by comparing the ratio of the characteristic strain of the source, $h_{c}(f)$, to the effective strain noise amplitude, $h_{n}(f)$, multiplied by the base-10 logarithm of the bandwidth, which is made simpler by the fact that these plots are generally shown in logscale. 
        
        In figure \ref{fig:sourceevol} we show a comparison of the chirping strain generated by IMRPhenomD for various source configurations.
        The cases displayed correspond to an array of coalescing BHBs with varying spin parameters, and whose masses and redshifts are chosen so that the characteristic strain lies in the frequency bands of PTAs and LISA.
        For each respective mass and redshift of the BHBs in the figure, we display the time to coalescence at different time slices in the inspiral.
        This allows for some insight into the timescale of inspiral versus the merger of various sources.
        
\subsection{Matched Filter and the Signal-to-Noise Ratio}
    Since the strain emitted from all but the strongest GW sources are below the instrument's noise fluctuations, we use matched filtering to maximize the sensitivity.
    Matched filtering can be used for many applications (e.g.,  parameter estimation, signal significance, etc.); in this case, we use it to find local maxima of the match between the signal and the data to maximize the SNR. 
    This method assumes that the noise in the signal is stationary and Gaussian, and that it is uncorrelated with the signal \citep{Maggiore2007}.
    
    In this work we use the typical optimal filter argument and arrive at,
    \begin{equation}{\label{eq:snrint2}}
        \rho^{2}=4\int_{0}^{\infty} \mathrm{d} f \frac{\left|\tilde{h}(f)\right|^{2}}{S_{n}(f)}=(\tilde{h}(f) \big| \tilde{h}(f)) .
    \end{equation}
    the most frequently used form of the optimal SNR, $\rho$.
    For a full derivation, see section 7.3 of \cite{Maggiore2007}.
    Using equations \ref{eq:instcharstrain} and \ref{eq:sourcecharstrain}, we can rewrite the SNR in equation \ref{eq:snrint2} as
    \begin{equation}{\label{eq:snrint3}}
        \rho^{2}=\int_{-\infty}^{\infty} \mathrm{d}(\ln f)\left[\frac{h_{\mathrm{c}}(f)}{h_{n}(f)}\right]^{2} ,
    \end{equation}
    which, as mentioned previously, allows for convenient visual estimation of SNR when the characteristic source strain and effective strain sensitivity of detectors are plotted together \citep{Flanagan1998} 
    
    \subsubsection{Monochromatic Sources}
        For monochromatic sources, the optimal filter is limited to the smallest bandwidth around the frequency of the binary.
        As mentioned in Section \S\ref{par:monosource}, instruments have a finite resolution limited by the observation time, $T_{\rm obs}$.
        Since we again use the average over solid angle and polarization, then using equation \ref{eq:snrint2} in tandem with \ref{eq:angavgstrain}, the squared SNR becomes, 
        \begin{equation}
            \rho^{2} = h^{2}_{0}(f_{0}) \int_{0}^{\infty} \mathrm{d} f \frac{\delta\left(f-f_{0}\right) \delta(0)}{S_{n}(f)}
        \end{equation}
        where $f_{0}$ is the frequency of the binary, $h_{0}$ is given in equation \ref{eq:monostrain}.
        
        To account for the finite observing time, we must replace the Dirac deltas with finite-time Dirac deltas,
        \begin{equation}
        \label{eq:delta}
            \delta(f)=\int_{-\infty}^{\infty} \mathrm{d}t~ e^{i 2 \pi f t} \rightarrow \int_{-T_{\rm obs} / 2}^{T_{\rm obs} / 2} \mathrm{d}t~ e^{i 2 \pi f t} ,
        \end{equation}
        which is allowed over the assumed window of equation \ref{eq:freqevol} \citep{Maggiore2007}. 
        In particular, from equation \ref{eq:delta}, $\delta(0) \rightarrow T_{\rm obs}$.
        
        Thus the squared SNR becomes, 
        \begin{equation}
            \label{eq:monoSNR}
            \rho^{2} = \frac{h^{2}_{0}(f_{0})~T_{\rm obs}}{S_{n}(f_{0})} =  \frac{h^{2}_{0}(f_{0})~N_{\mathrm{cycles}}}{h^2_{n}(f_{0})} = \left(\frac{h_{c}(f_0)}{h_{n}(f_{0})}\right)^2,
        \end{equation}
        where $N_{\mathrm{cycles}}$ is the number of waveform cycles that occur over the observation time, and the last equality serves to define $h_{c}$ for monochromatic binaries \citep{Maggiore2007}.
        Unlike in the chirping case where the waveform matching builds up over the changing frequency, the signal builds at the frequency of the source over the entire observation time.
        
    \begin{figure}[!htbp]
        \includegraphics[width = \textwidth]{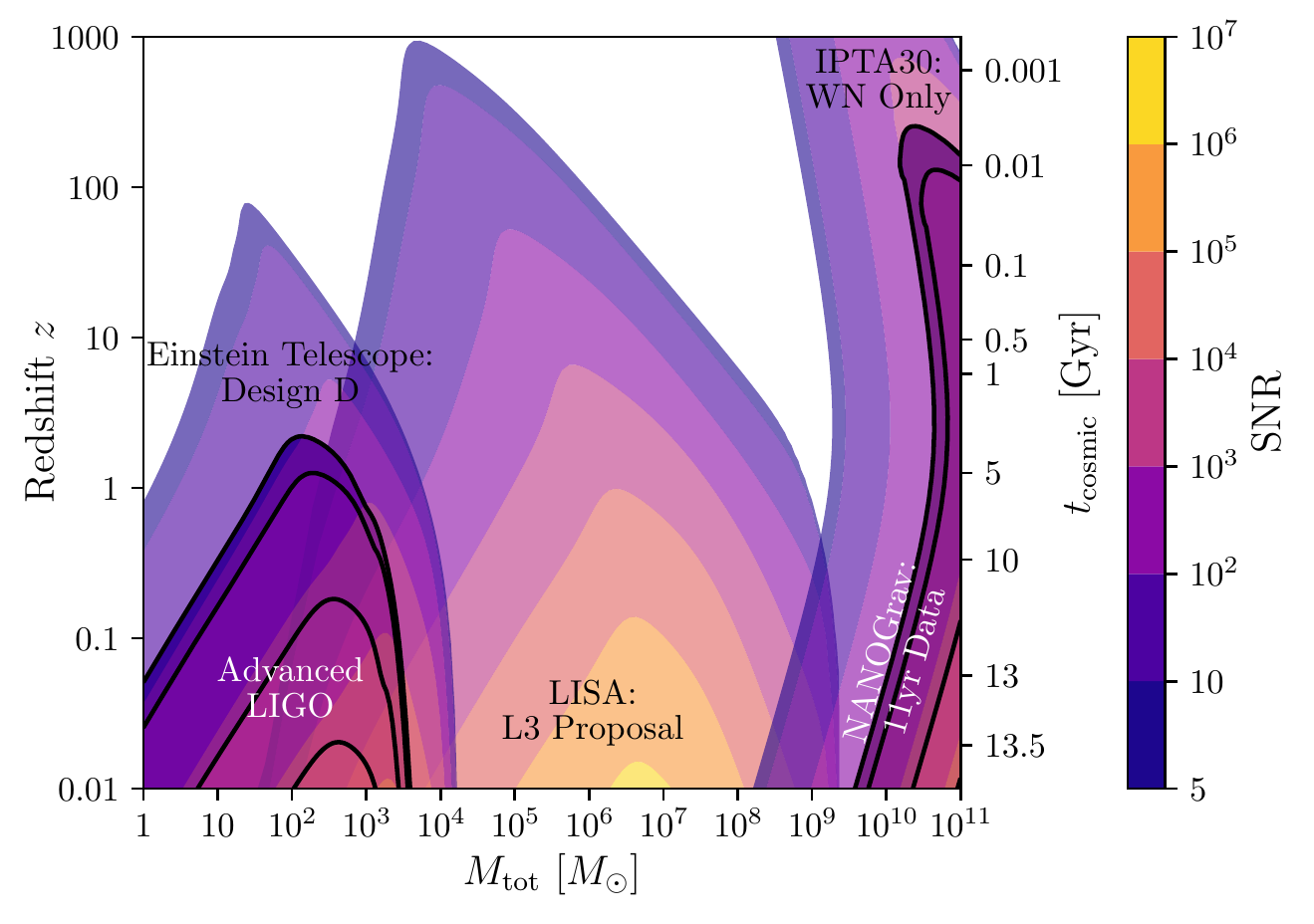}
        \caption{SNRs with respect to variations in source BHBs redshift and total source mass for ET design model D, aLIGO, LISA L3 proposal, IPTA30 WN only, and NANOGrav 11yr. All curves assume no stochastic foreground, an individual, non-spinning ($\chi_{i}=0$), equal mass ($q=1$) BHB as the source, and WMAP9 cosmological parameters. All detectors have complementary coverage to study the full range of black  hole masses at various stages of their evolution across the observable Universe. The particular regions of overlap between different detectors allow for observations of the same source class throughout the BHB's evolution.
        }
        \label{fig:fullSNR}
    \end{figure}    
    
        In figure \ref{fig:fullSNR}, we display the match-filtered SNR for the sensitivities of GW detectors across the entire gravitational-wave spectrum of coalescing BHBs.
        For figure \ref{fig:fullSNR} and for all relevant plots we use the flat $\Lambda$CDM, nine-year Wilkinson Microwave Anisotropy Prob (WMAP9) cosmology provided by \texttt{astropy} \citep{astropy2018,Hinshaw2013}.
        Using \texttt{gwent}, one can easily calculate the SNR for sensitivities of current and future GW detectors for any coalescing BHB parameter.
        Here, also using this formalism, we display the most relevant multiband overlap in parameter space in the literature, redshift and total source mass.
        Clearly, as instruments become more sensitive, the synergy between detector bands increases dramatically.

\section{PTA Sensitivity}
\label{sec:PTAsec}
PTAs search for sources of low-frequency GW signals by observing pulsar times-of-arrival (TOAs) and looking for deviations from models of the intrinsic pulsar TOAs.
Most of the analyses take place in residual space, where residuals are the predicted model TOAs subtracted from the observed TOAs.
In theory, one could detect GWs using individual pulsars \citep{Sazhin1978}. 
To attribute potential GW signatures in pulsar residuals to a signal and not some spurious noise source however, one needs to use correlations between pulsars.
By using correlations between different pulsars, global variances, like GW signals, can be extracted from the residuals.

Previous PTA sensitivity curve generators have often represented PTAs as a wedge with a sharp cutoff around $f = T_{\rm obs}^{-1}$ \citep{Thrane2013}.
In fact, PTAs do maintain some sensitivity at frequencies below this cutoff,
with the level being limited by pulsar timing model fitting \citep{Hazboun2019}.
One point of confusion with PTA sensitivity curve representation is the various ways in which they are presented. As pointed out in \cite{MooreTaylor2015}, one must be careful to not assume a similar response to different GW sources.

Here we focus on searches for GWs from supermassive black-hole binaries (SMBHBs).
Often these searches are grouped into two main categories: stochastic signals from a superposition of unresolved SMBHBs, and individually resolvable SMBHBs. 
This stochastic signal, referred to as the stochastic GW background (GWB), is common to all pulsars across the sky.
The GWB is hypothesized to follow a power law spectrum and induce a quadrupolar signature on the spatial correlations between each pairing of pulsars in the array, known as the Hellings-Downs (HD) correlation \citep{Hellings1983}.

Signals from individual SMBHBs do not induce an HD correlation between pulsars, but instead cause a correlation that depends on the source's location.
The strength of the correlation depends on the source's location and its amplitude as given by equation \ref{eq:monostrain}.
The formalism developed here focuses on these individual SMBHBs, not the GWB.

There are ongoing searches for both predicted types of GW signals, but we choose to focus on individually resolvable SMBHBs as they are directly comparable to BHB mergers across the GW spectrum.
For more information about searches for stochastic backgrounds, and the most recent NANOGrav PTA upper-limits, see \cite{nanoGWB2018}.
For the most recent results of the NANOGrav PTA search for individually resolvable SMBHBs, see \cite{nanoCW2018}.

\subsection{PTA Sensitivity Curves}
    Since we are interested in presenting sensitivities of PTAs in a similar way to the formalism used in ground and space-based interferometers, we outline PTA noise sources in terms of power spectral densities.
    In reality, PTAs operate first in the time domain in terms of TOAs and fitting individual pulsar noise properties.
    This is because unlike interferometers, PTAs do not have noise-only channels.
    Additionally, each pulsar can have unknown noise properties and sources.
    Thus, an amount of fitting is needed to characterize the TOAs of each pulsar \citep{nanoTiming2018}.
    Once a fit is generated for each pulsar, the modeled TOAs are subtracted from the observed TOAs, creating residuals that should only contain non-deterministic noise sources and signals (i.e.~, GWs).
    Our analysis treats PTAs in terms of these residuals, where the noise in individual pulsars simply contains some white and/or red noise that accounts for any remaining noise in the modeled TOAs.

    We use the sensitivity generation results of \cite{Hazboun2019} (via their code \href{https://pypi.org/project/hasasia/0.1.5/}{\texttt{hasasia}}) in constructing the characteristic strain plots for PTAs and in the SNR calculations.
    We reproduce the relevant equations for easy comparison between PTA sensitivity calculations and interferometers.

\subsection{PTA Power Spectral Density}
    The overall PSD of the residuals for a PTA is made up of the PSD from one or more stochastic backgrounds (SBs), $P_{\rm SB}(f)$, which is common to all $N_{\rm p}$ pulsars in the array, and the sum of all individual pulsar noises, $P_{i}(f)$ \citep{Thrane2013,Lam2018}: 
    \begin{equation}
        P_{n}(f) = P_{\rm SB}(f) + \sum_{i}^{N_{\rm p}}P_{i}(f) .
    \end{equation}
    Here we only consider power-law modeled SBs, which can be parametrized as
    \begin{equation}
        P_{\rm SB}(f) = \frac{A_{\rm SB}^{2}}{12 \pi^{2}}\left(\frac{f}{f_{\rm year}}\right)^{2\alpha}f^{-3} ,
    \end{equation}
    where $A_{\rm SB}$ is the strain amplitude at a frequency of 1 ${\rm yr^{-1}}$ (i.e., ${f_{\rm year}}$) and $\alpha$ is the spectral index of the characteristic strain. 
    For a SB dominated by SMBHBs (i.e., the GWB), $\alpha = -2/3$ is assumed \citep{Jenet2006}. 
    We assume a conservative amplitude $A_{\rm GWB} = 4\times10^{-16}$ found as the median value of SMBHBs merger rate estimates in \cite{2016Sesana}.
    If we include the GWB in our framework, it appears as the dominant ``instrument" noise at lower frequencies.
    
    Each pulsar's PSD,
    \begin{equation}
        \label{eq:PulsarPSD}
        P_{i}(f)  =  P_{{\rm RN},i}(f) + P_{{\rm WN},i} ,
    \end{equation}
    is made up of two components: 
    the pulsar red noise (RN),
    \begin{equation}
        P_{{\rm RN},i}(f) = A_{{\rm RN},i} \left(\frac{f}{f_{\rm year}}\right)^{-\gamma_{i}} , \hspace{5mm} \gamma_{i} > 0,
    \end{equation}
    where $A_{{\rm RN},i}$ is the amplitude of the pulsar red noise and $\gamma_{i}$ is the red noise power spectral index, 
    and the pulsar white noise (WN),
    \begin{equation}
        \label{eq:PSDWi}
        P_{{\rm WN},i} = 2\,\frac{\sigma_{i}^{2}}{c_{i}},
    \end{equation}
    which is determined by the observing cadence $c_{i}$ ($c_{i}$ is also written as $1/\Delta t_{i}$ in the literature) and $\sigma_{i}$, the white-noise rms of each pulsar.
    
    To implement these noise PSDs appropriately, one needs to carefully account for the power removed by fitting the timing model.
    This is represented in the diagonal inverse-noise-weighted transmission function,
    \begin{equation}
        \label{eq:Hazbouneq21}
        \mathcal{N}_{i}^{-1}(f)=\mathcal{T}_{i}(f) / P_{i}(f) ,
    \end{equation}
    where $\mathcal{T}_{i}(f)$ is the pulsar's transmission function, and $P_{i}(f)$ is the white noise PSD in equation \ref{eq:PSDWi} \citep{Hazboun2019}.
    $\mathcal{T}_{i}(f)$ characterizes the diminished sensitivity to frequencies removed by fitting a timing model, and is treated in detail in \cite{Hazboun2019}, but not replicated here for brevity.
    For the inclusion of red and white noise in the pulsar (i.e., using \ref{eq:PulsarPSD}), equation \ref{eq:Hazbouneq21} becomes approximate and the noise covariance matrix becomes non-diagonal \citep{Hazboun2019}.
    
    \subsubsection{The Response Function}{\label{subsubsection:PTAResponse}}
    The pulsar timing residual response function is given by
    \begin{equation}
        \label{eq:PTAresponsefunction}
        R^{+, \times}(f, \widehat{\Omega}, \psi) = \frac{1}{i 2 \pi f}\left(1-e^{-i 2 \pi f D(1+\widehat{\Omega} \cdot \hat{p}) / c}\right)\sum_{A=+,\times}F^{A}(\widehat{\Omega},\psi) ,
    \end{equation}
    where
    \begin{equation}
        F^{A}(\widehat{\Omega},\psi) \equiv \epsilon_{i j}^{A}(\hat{\Omega},\psi) \frac{\hat{p}^{i} \hat{p}^{j}}{2\left(1+\hat{\Omega} \cdot \hat{p}\right)} ,
    \end{equation}
    $\hat{p}$ is the unit vector from the solar system barycenter to the pulsar, $D$ is the distance to the pulsar, $\epsilon_{i j}^{A}(\hat{\Omega},\psi)$ are given in equation \ref{eq:PolarizationTensors} and $i,j = x,y,z$ are the spatial indices \citep{Anholm2009}.
    Note the factor of $\left(1-\mathrm{exp}(-i 2 \pi f D(1+\widehat{\Omega} \cdot \hat{p})/c)\right)$, which comes from the redshifting of signals from a pulsar and is the metric perturbation at the Earth (first term) and the pulsar (second term). 
    We note that the pulsar term is averaged over in \cite{Hazboun2019} and produces a factor of two increase in sensitivity over simply taking the Earth term only. 
    The factor of $(1/i2\pi f )$ comes from transforming the redshift in pulse arrival to the residuals which are typically used in pulsar timing \citep{Hazboun2019}. For a more in-depth discussion, see section 2 in \cite{Anholm2009}.
    
    The sky and polarization averaged response function in equation \ref{eq:angavgstrain} for PTAs is then given by
    \begin{equation}
        \mathcal{R}(f) = \frac{1}{12\pi^{2}f^{2}} ,
    \end{equation}
    where the detector response functions in equation \ref{eq:angavgstrain} come from equation \ref{eq:PTAresponsefunction} to account for the Earth and pulsar terms.
    
    To assemble a full PSD one sums the individual pulsar strain-noise PSD,
    \begin{equation}
        S_{I}(f) \equiv \frac{1}{\mathcal{N}_{I}^{-1}(f) \mathcal{R}(f)},
    \end{equation}
    scaled by the individual pulsar's fractional observation time to get the effective noise PSD,
    \begin{equation}
        S_{\mathrm{eff}}(f) \equiv\left(\frac{4}{5} \sum_{I} \frac{T_{I}}{T_{\mathrm{obs}}} \frac{1}{S_{I}(f)}\right)^{-1} .
    \end{equation}
    The factor of $4/5$ again comes from averaging over inclination.
    We assume that $S_{n}(f)=S_{\mathrm{eff}}(f)$ when calculating noise curves and SNRs \citep{Hazboun2019}.
    
\subsection{Sampled PTAs}
    \label{subsec:sampledPTAs}
\begin{figure}[!htbp]
        \includegraphics[width=\textwidth]{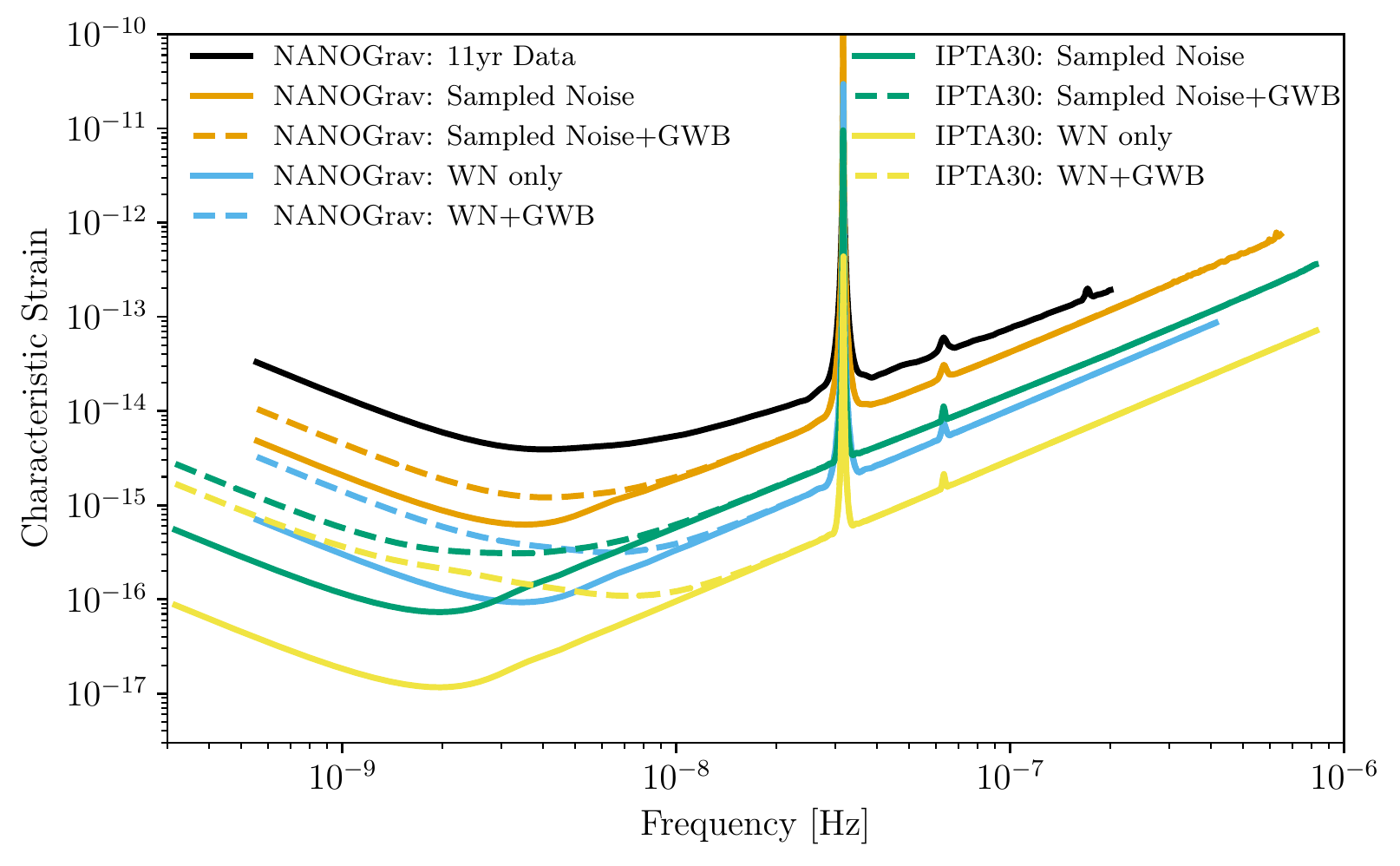}
    
        \caption{Sensitivities of various PTA designs with noise parameters noted in table \ref{tab:PTAparams}.
        Dashed curves use the same values as the solid curves, but with an added GWB of $4\times10^{-16}$.
        The NANOGrav 11yr data sensitivity curve uses the real TOAs and fitting parameters given in \cite{nanoTiming2018}.
        The WN only models have identical fixed WN levels for the number of pulsars in the array observed over the total respective timespan.
        The difference between the IPTA30 and NANOGrav WN only models is that the IPTA30 model includes an increase in the number of pulsars, cadence, and total observation time.
        The sampled noise cases either assign each pulsar with corresponding values in \cite{nanoTiming2018} for the NANOGrav cases, or draw values from uniform distributions given in table \ref{tab:PTAparams} for the IPTA30 cases.
        The specific makeup of the sampled noise cases are more rigorously defined in \S\ref{subsec:sampledPTAs} and in table \ref{tab:PTAparams}.
        It is clear that the simplified frameworks employed here tend to underestimate the noise by not including some noise sources, leading to an increase in the sensitivity of the PTA.}
        \label{fig:hasasiaPTASensitivity}
    \end{figure}

    Using the robust methods presented in \cite{Hazboun2019} in this formalism allows us to compare future detectors to current ones using improved methods to previous PTA SNR calculations. 
    Additionally, using the framework of \cite{Hazboun2019} we are able to construct sensitivities for PTAs with more realistic noise parameters, and to use real data directly. 
    In figure \ref{fig:hasasiaPTASensitivity}, we show the characteristic strain for five different NANOGrav models and four estimated International PTA (IPTA) models for an IPTA in the 2030s similar to those in \cite{Chen2019} and \cite{Taylor2016}. 
    
    All of the generated models improve upon the sensitivity of the NANOGrav 11yr data curve in figure \ref{fig:hasasiaPTASensitivity}.
    The NANOGrav 11yr data sensitivity curve uses the real TOAs and fitting parameters given in \cite{nanoTiming2018} to give a comparison to simulated plots to real data in \texttt{hasasia}.

    The WN only models in figure \ref{fig:hasasiaPTASensitivity} have identical fixed WN levels for the number of pulsars in the array (see table \ref{tab:PTAparams} for exact values).
    The difference between the IPTA30 and NANOGrav WN only models is that the IPTA30 model includes an increase in the number of pulsars, an increase in cadence, and a total observation time of 20 years instead of the NANOGrav model's 11 years.
    This is reflected by the IPTA30 WN only model's almost order of magnitude lower sensitivity curve in figure \ref{fig:hasasiaPTASensitivity} when compared to the NANOGrav WN only model.
    When we take the same two WN only models and add a GWB, the sensitivity of both models is reduced at frequencies below $f_{\rm year}$.
    It is expected that adding a GWB as the only source of red noise in all pulsars would cause the low frequency sensitivity to decrease and approach similar values for PTAs with different WN levels.
    To fairly and accurately characterize PTA sensitivities, we should be considering the noise properties of the entire array of pulsars instead of the traditional simplistic, WN only models. 
    
    To simulate more realistic PTAs, labeled as ``Sampled Noise" in figure \ref{fig:hasasiaPTASensitivity}, we use current WN and RN noise parameters along with pulsar sky locations, observation times, and cadences from \cite{nanoTiming2018}.
    Each pulsar that makes up the sampled NANOGrav curves is assigned an RMS and a RN.
    Thus, for these PTAs, the overall RMS from table 2 in \cite{nanoTiming2018} is used as the WN RMS, and the RN levels are set to the maximum likelihood values found in the analysis of \cite{nanoGWB2018}.
    For the NANOGrav curves, we use only the 34 pulsars used in \cite{nanoGWB2018} with their corresponding locations, and observation and noise parameters.
    For the IPTA30 sampled noise models, we use the individual NANOGrav 11 year parameters to create a distribution of values from which to draw each pulsar's location and RN parameters.
    The WN parameters for the sampled IPTA30 models use a uniform distribution between the WN-only model's  $\sigma_{\mathrm{RMS}}$ and ten times that value.
    The per pulsar cadence is likewise drawn from a uniform distribution of observing between once and four times a month.
    The distribution ranges for all sampled models are given in table \ref{tab:PTAparams}.
    Drawing from real pulsar noise data gives us a 
    reasonable representation of sensitivities that a future PTA can obtain without needing to simulate a full dataset.
    
    \begin{table*}[t]
    \fl
    \caption{PTA component noise values used throughout this work. 
    Brackets indicate the ranges for the relevant noise values. 
    Ranges for the sampled NANOGrav parameters and IPTA30 red noise parameters correspond to the minimum and maximum noise values in the NANOGrav 11yr dataset \citep{nanoTiming2018}.
    The IPTA30 values correspond to an hypothetical IPTA in the 2030s similar to those in \cite{Chen2019} and \cite{Taylor2016}.
    $\sigma_{\mathrm{RMS}}$ is the white-noise rms of each pulsar, the cadence is the number of observations per pulsar per month, $T_{\mathrm{obs}}$ is the total observation time. $A_{\mathrm{GWB}}$ is the amplitude of the stochastic GW background at $f = 1/\mathrm{year}$, $A_{\mathrm{RN}}$ is the amplitude of the pulsar red noise at $f = 1/\mathrm{year}$, and $\gamma_{\mathrm{RN}}$ is the red noise power spectral index.
    We note that we use the overall RMS from \cite{nanoTiming2018} as the bounds on the sampled NANOGrav $\sigma_{\mathrm{RMS}}$.
    Since the overall RMS may include RN, this may inflate the WN levels, but since we do not include other sources of WN present in real TOAs (e.g. EQUAD, EFAC, and ECORR), we expect this choice to produce little effect on the resulting sensitivity levels.
    }
    \vspace{2mm}
    \resizebox{\columnwidth}{!}{%
    \begin{tabular}{@{} llllllll @{}}
        \hline\hline
            Model & $\sigma_{\mathrm{RMS}}$& cadence & $T_{\mathrm{obs}}$ & $N_{\mathrm{p}}$ & $A_{\mathrm{GWB}}$ & $\mathrm{log}_{10}(A_{\mathrm{RN}})$ & $\gamma_{\mathrm{RN}}$\\ 
            & [ns] & [$\mathrm{month}^{-1}$] & [yr] & & $\times 10^{-16}$ & & \\
            \hline

            NANOGrav: WN Only & 100 & 2 & 11.4 & 34 & --- & --- & --- \\
            NANOGrav: WN + GWB & 100 & 2 & 11.4 & 34 & 4 & --- & ---  \\
            NANOGrav: Sampled Noise & [108,3650] & [1,4] & [4,11.4] & 34 & --- & [-19.9,-12.2] & [-1.85,6.24] \\
            NANOGrav: Sampled Noise + GWB & [108,3650] & [1,4] & [4,11.4] & 34 & 4 & [-19.9,-12.2] & [-1.85,6.24] \\
            IPTA30: WN Only & 100 & 4 & 20 & 200 & --- & --- & --- \\
            IPTA30: WN + GWB & 100 & 4 & 20 & 200 & 4 & --- & --- \\
            IPTA30: Sampled Noise & [100,1000] & [1,4] & 20 & 200 & --- & [-19.9,-12.2] & [-1.85,6.24] \\
            IPTA30: Sampled Noise + GWB & [100,1000] & [1,4] & 20 & 200 & 4 & [-19.9,-12.2] & [-1.85,6.24] \\
            \hline
    \end{tabular}
    \label{tab:PTAparams}%
    }
\end{table*}

    As can be seen in figure \ref{fig:hasasiaPTASensitivity}, the result of this sampled PTA generation still produces a discrepancy of about a factor of three relative to noise levels present in real TOAs/residuals.
    To fully simulate a PTA's sensitivity, one would need to account for additional noise sources, the origins of which are not yet entirely clear \citep{Lam2017}.
    In order to utilize \texttt{gwent}'s unique ability to produce SNRs to BHB sources for variations in instrument parameters, we cannot realistically simulate full PTA datasets that attempt to include the unknown sources of noise.

\subsection{Methods of Sensitivity Representation}
    \label{subsec:otherPTAs}
\begin{figure}[!htbp]
        \includegraphics[width = \textwidth]{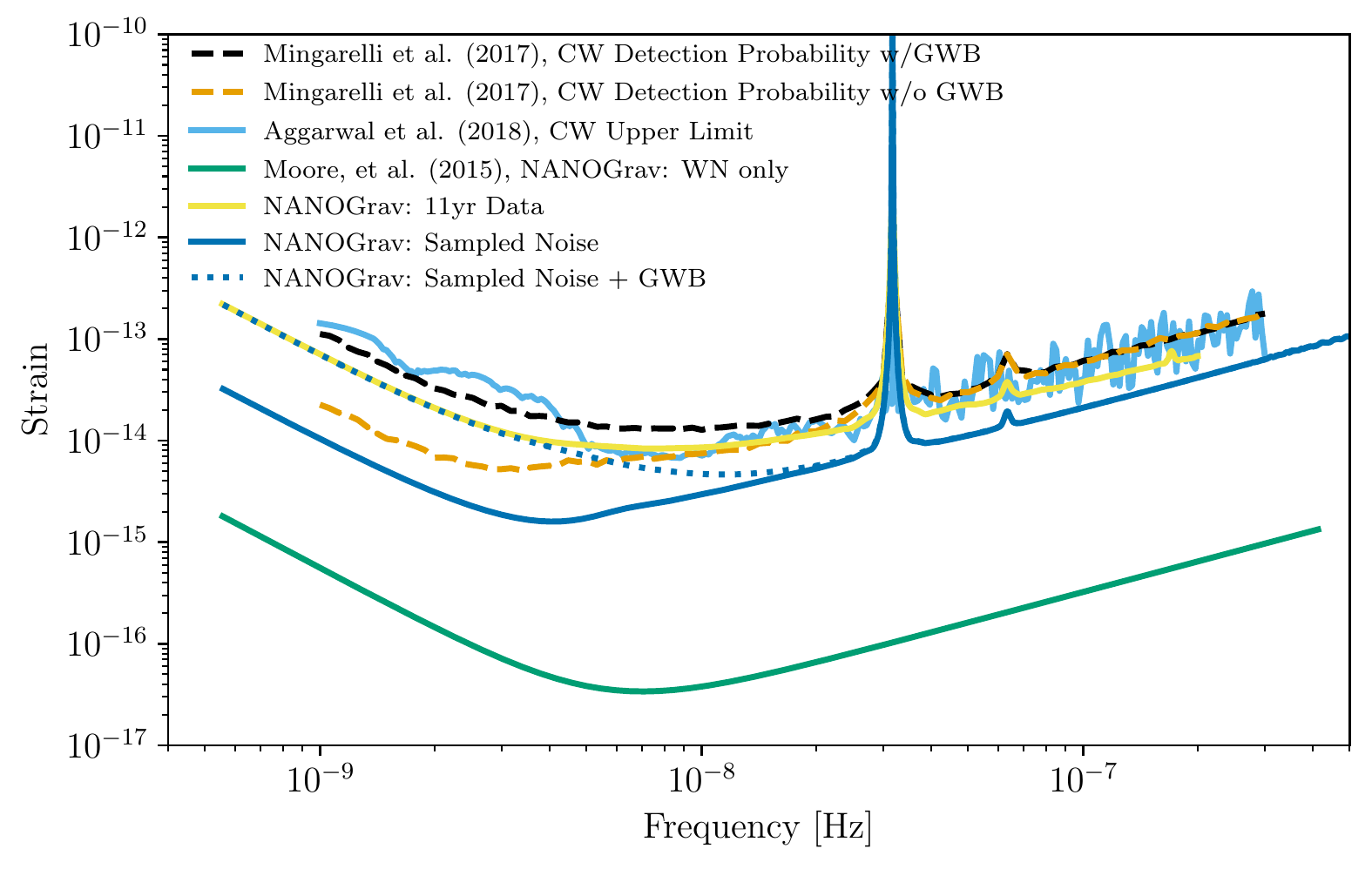}
        \caption{
        Sensitivities of various PTA designs discussed in \S\ref{subsec:otherPTAs}. 
        The NANOGrav 11 year data curve and the NANOGrav sampled noise curve use the same parameters as in figure \ref{fig:hasasiaPTASensitivity}.
        For the NANOGrav sampled noise curve that includes a GWB, we use a background amplitude of $A_{\rm GWB} = 2.0 \times 10^{-15}$ in order to better match the real data sensitivity.
        In order to present the curves produced by our formalism with those mentioned in section \S\ref{subsec:otherPTAs}, we plot the strain of each PTA to a source that produces an SNR of 3 via solving equation \ref{eq:monoSNR} for $h_{0}$.
        The \cite{Mingarelli2017} curves use the 95\% detection probability with a false alarm probability of $10^{-4}$, a 15 yr observation time, and for the GWB case, an unsubtracted GWB amplitude of $4\times 10^{-16}$.
        The WN only, scaled detection sensitivity curve derived from \cite{MooreTaylor2015} has a similar spectral shape to the other PTAs in this figure, but has a distinctly more sensitive detection strain for the same choice of parameters. 
        }
        \label{fig:otherPTASensitivity}
    \end{figure}

    As mentioned earlier, one must be careful when comparing various PTA sensitivity curves.
    We again note that one cannot construct a PTA sensitivity curve agnostic of the source.
    For this reason, one should view PTA curves for deterministic sources as different from curves for stochastic sources.
    For example, the wedge-shaped sensitivity curves in  \cite{Thrane2013}, \cite{Hobbs2011}, and \cite{MooreCole2015} are often used to represent PTA sensitivities.
    Since these curves represent PTA's sensitivities to the GWB, one should not use them in the context of individually resolvable SMBHBs.
    With that in mind, we briefly examine three typical variations of PTA sensitivities to continuous waves (CWs, i.e., monochromatic GW sources) from SMBHBs and compare them to our methods.
    We display an example of each of the discussed models in figure \ref{fig:otherPTASensitivity}.
    
    In figure \ref{fig:otherPTASensitivity}, we also present three curves used in this study to highlight the differences between the various methods of assessing PTA sensitivities to CWs.
    The NANOGrav 11 year data curve and the NANOGrav sampled noise curve use the same parameters in figure \ref{fig:hasasiaPTASensitivity}.
    For the NANOGrav sampled noise curve that includes a GWB, we use a background amplitude of $A_{\rm GWB} = 2.0 \times 10^{-15}$.
    This singular change is more in line with current NANOGrav upper-limits on the GWB, and reduces the difference in sensitivity between the simulated curves and real data to within a factor of around $1.5$.
    Because characteristic strain curves, as seen in figure \ref{fig:hasasiaPTASensitivity}, do not directly relate to the source strain amplitude, we plot the strain of each PTA to a source that produces an SNR of 3 via solving equation \ref{eq:monoSNR} for $h_{0}$.
    This allows us to directly compare to other methods in the literature typically used for assessing detectability of CW sources.
    
    \cite{nanoCW2018} used Bayesian inference in addition to frequentist techniques to search for GWs from individual SMBHBs in circular orbits in their 11-year dataset.
    For the Bayesian GW analyses, the authors used fixed individual pulsar white noise parameters based on previous noise analyses, and sampled the red noise parameter space.
    To extract the common red noise (i.e., the CW) from the intrinsic pulsar red noise, the authors searched over both simultaneously. 
    For the frequentist analysis, the authors used the $\mathcal{F}_{p}$-statistic developed in \cite{Ellis2012}, which maximizes the log of the likelihood ratio of all pulsar-dependent contributions. 
    The $\mathcal{F}_{p}$-statistic is the easiest to relate to the analysis techniques presented here because it is directly related to the optimal SNR
    \begin{equation}
        \label{eq:Fpstat}
            \left\langle 2 \mathcal{F}_{p}\right\rangle = 2 M+(\tilde{\mathbf{s}} | \tilde{\mathbf{s}}) = 2 M+\rho^{2},
    \end{equation}
    where $\tilde{\mathbf{s}}$ are the induced pulsar timing residuals, which can be related to the source signal, $\rho$ is the optimal SNR, and $2M$ are the pulsar-dependent amplitude parameters; see \cite{Ellis2012} for more detailed definitions.
    To use the $\mathcal{F}_{p}$-statistic to produce sensitivity curves similar to the work here, one could calculate the $\mathcal{F}_{p}$-statistic corresponding to a minimum detectable SNR in each frequency bin.

    To get the SNR to compare directly to the formalism used here using the $\mathcal{F}_{p}$-statistic, one would need to either simulate, or use real-pulsar data to populate, a PTA including pulsars with particular noise data over some observation time-length.
    Then, one would need to inject a signal corresponding to a particular source into the PTA, including its response based on its sky location.
    Once the signal is injected, one would then compute the $\mathcal{F}_{p}$-statistic and recover the SNR from equation \ref{eq:Fpstat}.
    
    The issue with this method is that one needs to assign a sky-location to the source.
    This is because the response in the detector is heavily dependent on the arrangement of pulsars in the PTA.
    A region of the sky with fewer pulsars leads to insensitivities to sources in that particular region \citep{nanoCW2018}.
    This presents a problem as well as an opportunity, because PTA's sky sensitivities evolve as more pulsars are added, especially in underrepresented regions.
    
    Since neither their Bayesian nor their frequentist searches showed significant evidence of a CW, \cite{nanoCW2018} computed the sky-averaged $95\%$ upper-limit of the GW strain.
    Because these upper-limit curves are therefore not directly related to the sensitivity calculations as we have described it, one cannot use upper-limit curves for the SNR calculations we wish to perform.
    This also means that the use of EPTA upper-limit curves in \cite{Rosado2016} to produce horizon distance estimates is inconsistent, although one would need to recalculate those results using a realization of the EPTA effective noise PSD to determine how large a difference this would make.
    
    In \cite{Mingarelli2017}, the authors use the Two Micron All Sky Survey (2MASS) \citep{Skrutskie2006} in tandem with simulated galaxy merger rates from the Illustris cosmological simulation project \citep{Rodriguez-Gomez2015,Illustris2014} to populate the local universe (all GW sources are placed at distances $<~225$ Mpc).
    After simulating multiple realizations of the Universe, the authors then use the $\mathcal{F}_{p}$-statistic to compute detection probability curves at varying False Alarm Probabilities (FAP), or the probability that a measured $\mathcal{F}_{p}$ exceeds an $\mathcal{F}_{p}$ with no signal present \citep{Ellis2012}.
    The authors then produce sky-averaged strain sensitivity curves and strain sensitivity sky-maps at the particular locations of the 2MASS galaxies.
    
    Because the curves in \cite{Mingarelli2017} correspond to particular FAPs, they are detection threshold sensitivity curves. 
    Each curve corresponds to the particular detection confidence threshold corresponding to each FAP. 
    Therefore, one should not use the results of \cite{Mingarelli2017} directly as a sensitivity curve within our formalism.
    One could, however, use their results as another comparison of the detection significance and strength of the corresponding signal.
    
    In \cite{MooreTaylor2015}, the authors' use both frequentist and Bayesian methods to generate PTA sensitivity curves to both a monochromatic GW source (a CW) and a power-law stochastic background (the GWB).
    The author's produce scaling laws for the sensitivity to a CW.
    The frequentist method uses a similar formalism to that of \cite{Hazboun2019}, but assumes only white, Gaussian, and uncorrelated noise between all pulsars in the array.
    This allows for a great simplification of the PTA's response to a CW.
    Likewise, they only consider an optimally oriented source ($\iota=\psi=0$) and average over the sky position angles to get the scaling for the SNR of a monochromatic source.
    By choosing a detection threshold SNR, one can then invert their result to find the PTA's sensitivity to a CW.
    This is useful for scaling relations for the sensitivity of a simplified PTA to a CW.
    Since our model is intended to extend to chirping sources, and we wish to more accurately represent PTAs, we do not use their relations.
    
\subsection{Sensitivity Changes to Various Designs}
\begin{figure}[!htbp]
        \includegraphics[width=\textwidth]{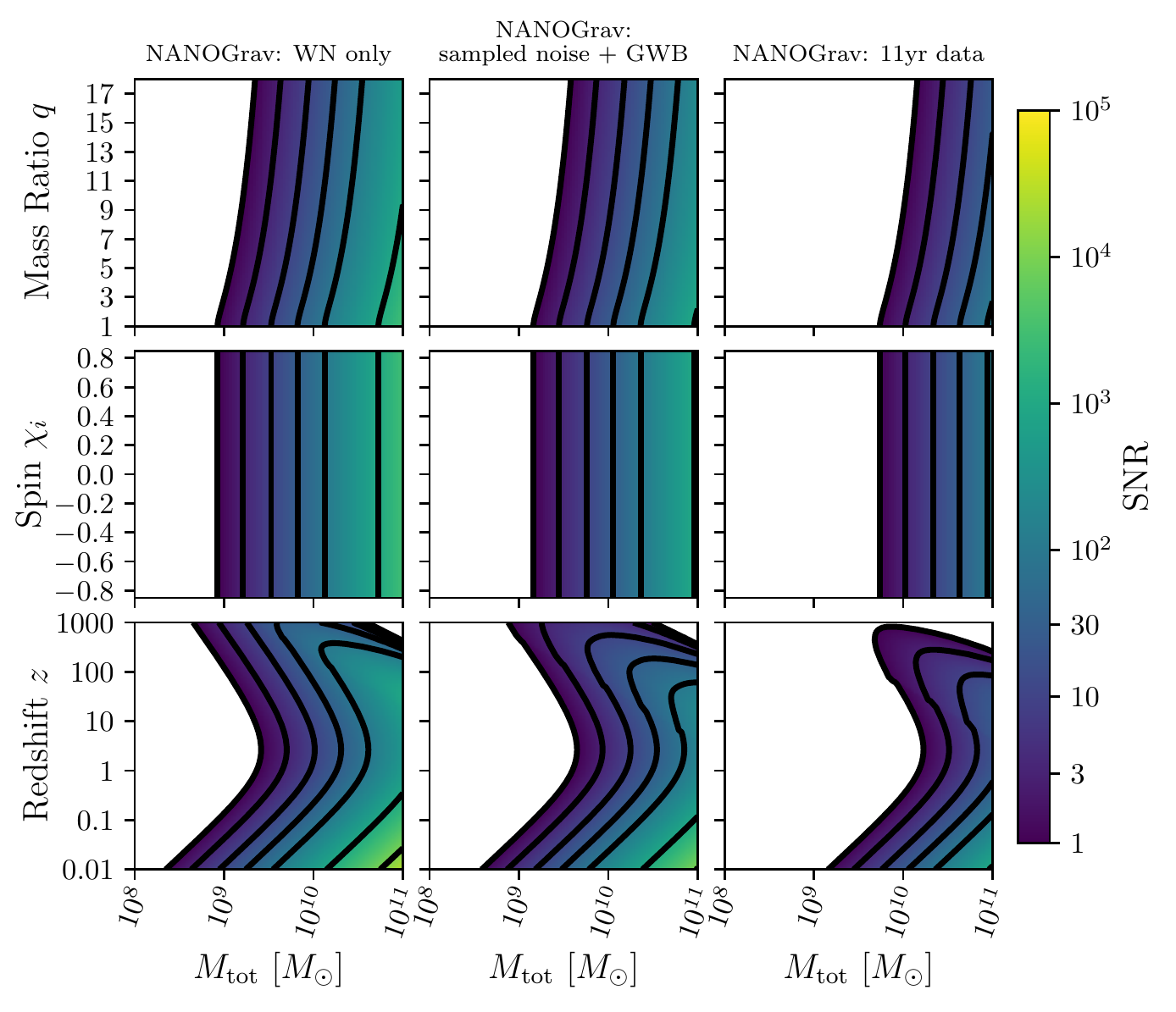}
        \caption{SNRs for changes to a fiducial, non-spinning, equal-mass BHB at a redshift of $z=0.1$ ($\chi_{i}=0, q=1$) with respect to PTAs: NANOGrav: WN only (Left Column), NANOGrav: sampled noise + GWB (Middle Column), and NANOGrav: 11yr data (Right Column). Solid black contours correspond to the levels marked on the colorbar. \textbf{Top Row:} mass ratio vs total source mass, \textbf{Middle Row:} spins of both BHs $\chi_{i}=\chi_{1}=\chi_{2}$ vs total source mass, \textbf{Bottom Row:} redshift vs total source mass. 
        }
        \label{fig:NANOGravmodelComp}
    \end{figure}

    In figure \ref{fig:NANOGravmodelComp}, we show the SNR versus total source mass for variations of a non-spinning, equal-mass GW source at a redshift of $z=0.1$ (i.e., two of those three parameters are held fixed while the third is varied).
    We simultaneously present the same changes for three NANOGrav curves in figure \ref{fig:hasasiaPTASensitivity}.
    Not surprisingly, the full range of sensitive spins, mass ratios, redshifts, and masses decrease as we continue to add noise sources.
    
    In other words, the further away one gets from the WN only model (i.e., the optimal PTA scenario of identical pulsars with only stochastic noise), the more the ranges of detectable source distance and total mass decrease. 
    
    {}

    For higher masses, there can be some change in the waveform close to merger with information about the spins of the individual binaries.
    However, based on figure \ref{fig:sourceevol}, PTAs are not expected to be sensitive to spin effects on the waveform amplitude.
    Even for the most massive and nearby sources that might be considered astrophysically relevant, the difference between a spinning and a non-spinning system is minimal over the frequencies which a PTA could observe the BHB. 
    
    As one can see, the effect of varying mass ratio has the same level of impact on the SNR for any time in the BHB's evolution.
    In all three detectors, there is a significantly larger total mass parameter space observable for equal mass binaries than for binaries with large mass ratios.

    \begin{figure}[!htbp]
        \includegraphics[width=\textwidth]{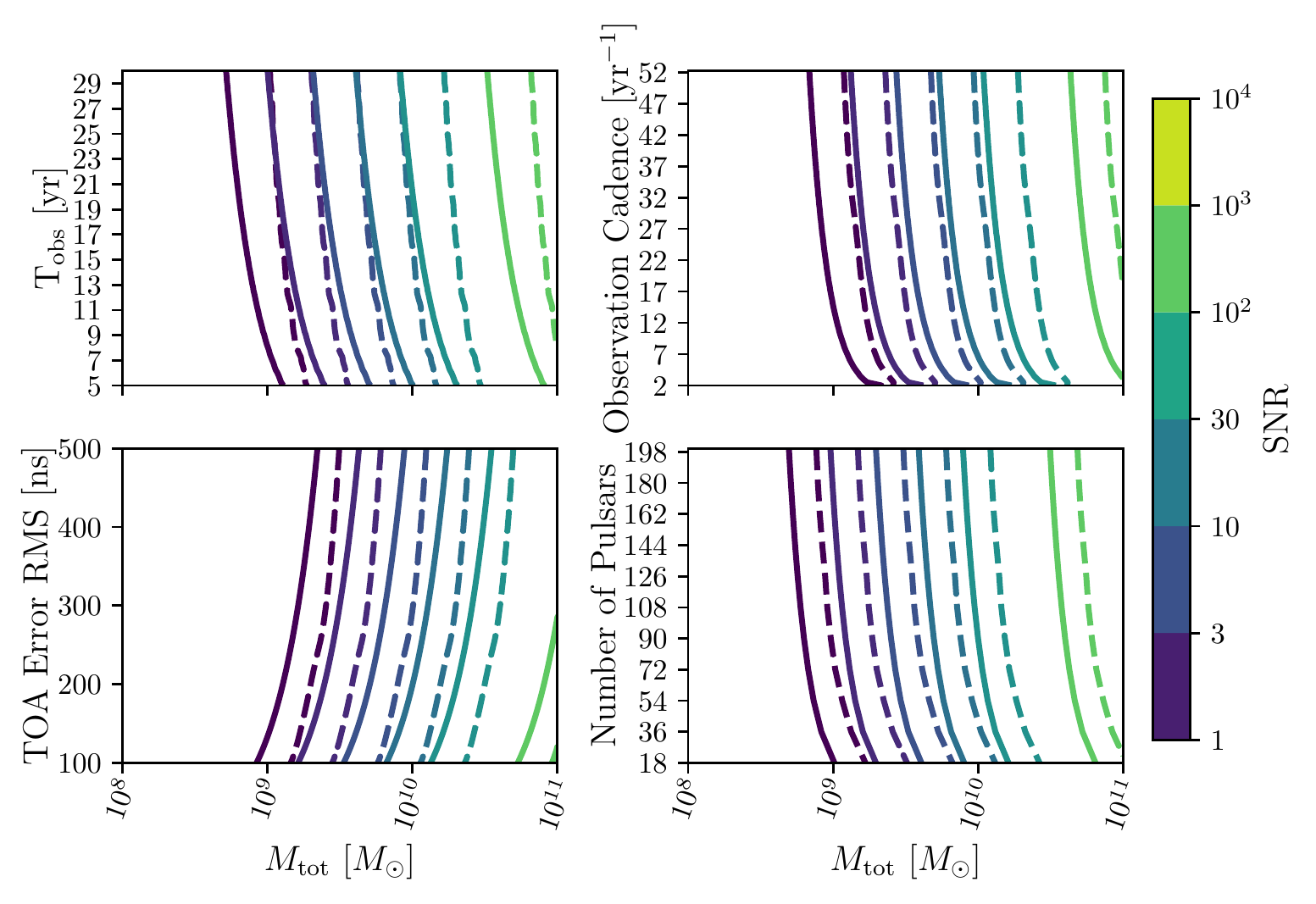}
        \caption{
        SNRs of non-spinning, equal mass BHB sources at redshift $z=0.1$ observed by variations on the fiducial NANOGrav: WN Only (solid lines) and NANOGrav: sampled noise + GWB (dashed lines) as a function of total mass. Each parameter is changed in isolation while the other PTA parameters are fixed to their fiducial values (see table \ref{tab:PTAparams}). \textbf{Upper Left:} observing time, \textbf{Upper Right:} observing cadence, \textbf{Lower Left:} pulsar timing noise RMS, \textbf{Lower Right:} number of pulsars.}
        \label{fig:NANOGravparamComp}
    \end{figure}{}
    
    We also use this formalism's unique ability to vary instrument parameters, to produce SNRs that show examples of how particular noise parameters affect PTA sensitivities to specific GW sources. 
    In figure \ref{fig:NANOGravparamComp} we take the fiducial values for the NANOGrav WN only and NANOGrav sampled noise + GWB PTAs (see table \ref{tab:PTAparams}), then vary the observation time, cadence, WN RMS, and the number of pulsars, respectively. 
    We then calculate the SNR for variations to these parameters versus the total source mass for a non-spinning, equal-mass SMBHB at a redshift of $z=0.1$ on the other axis.
    This allows us to explore the change in SNR over the full variable range for each of the above parameters.
    We find that the SNRs behave as one would expect: as observation time, cadence, and the number of pulsars increases, and as the timing RMS decreases, the PTA gains sensitivity overall, and consequently a larger range of total mass becomes detectable. 

    We find that increasing the cadence extends PTA sensitivity to higher frequencies and, more importantly in this context, lowers the overall WN.
    As the PTA's higher frequency coverage increases, and the optimal frequency does not change.
    The SNR varies with cadence, $c$, as $\sqrt{c}$.
    The WN is also lowered by decreasing the timing error RMS $\sigma_{\mathrm{RMS}}$, which is represented by the SNR scaling as $1/\sigma_{\mathrm{RMS}}$.
    When adding multiple pulsars with similar $T_{\mathrm{obs}},~\sigma_{\mathrm{RMS}}$, and $c$ to a WN only PTA, the SNR scales with the number of pulsars $N_{\mathrm{p}}$ as $\sqrt{N_{\mathrm{p}}}$.
    
    Since the SMBHBs with masses smaller than $\sim10^{10} \mathrm{M}_{\odot}$ tend to be monochromatic sources in the PTA band, the SNR only builds as $\sqrt{T_{\mathrm{obs}}}$ for those sources.
    Increasing $T_{\mathrm{obs}}$ extends sensitivity to lower frequencies, and also shifts the optimal frequency of the detector to lower frequencies.
    The shift in optimal frequency adds an additional factor to the typical scaling of $\sqrt{T_{\mathrm{obs}}}$ with SNR for a WN only PTA.
    
    It is clear that, in the case of WN only pulsars, when increasing the observation time, cadence, or number of pulsars, there is a point of diminishing returns because of their effect on SNR.
    The only parameter that does not follow this trend is the WN timing error.
    In the case of using sampled noise, the SNRs follow similar trends as the WN only model, but with around an order of magnitude less sensitivity overall.
    This is to be expected based on a similar factor in sensitivity decrease between the WN only model and sampled noise models for a  NANOGrav-esque PTA seen in figure \ref{fig:hasasiaPTASensitivity}.
    
    The biggest difference in the shape of the sampled noise curves is their smoothness when varying some parameters.
    When changing parameters in the sampled noise case, the SNR appears to have a non-smooth increase when comparing to the WN only case.
    we attribute this to the fact that when varying each parameter, the entire array is assigned the new sampled value.
    We find that simply changing one factor only has a moderate effect on the overall SNR, but we emphasize that this conclusion is highly dependent on our underlying assumption of a homogeneous array.
    In reality, when increasing the number of pulsars, the SNR of a source in the direction of a newly added pulsar could drastically increase.
    This is because the non-sky averaged sensitivity is highly dependent on the relative locations of the source and pulsars \citep{nanoCW2018}.
    
    Another interesting finding using the capabilities of this formalism is that the SNR dips below the threshold for detection, then comes back into the detectable range at higher redshifts for certain sources.
    This effect, seen in figure \ref{fig:NANOGravmodelComp}, was also noted in \cite{Rosado2016}.
    Because of the increase in SNR at high redshifts, PTAs could potentially detect sources in the very early Universe, were they to exist.
    Typically these high mass and high redshift sources have some orbital evolution during the observation time (i.e., they don't obey equation \ref{eq:freqevol}).  
    We emphasize that much of the domain of these plots covers regions of parameter space (e.g., $10^{11}\,M_{\odot}$ BHs at $z=1000$) that should not exist within our current understanding of cosmological structure formation; these plots are strictly intended to illustrate the potentiality of a PTA measurement, rather than our expectation based on current theory.

\section{Space Based Detector Sensitivity}
\label{sec:LISAsec}
In this section we discuss space-based detectors in the context of LISA, a detector consisting of three satellites in a triangular constellation connected by two-way laser links \citep{2017AmaroSeoane}. 
The LISA frequency band spans from $\sim$10~$\mu$Hz to $\sim$1~Hz (See figure \ref{fig:LISASensitivity}). 
To get to the lower frequency domain, the proposed inter-spacecraft separation of LISA will be on the order of 2.5 million km. 
The LISA three satellite constellation will trail Earth between 50 and 65 million km in a heliocentric orbit and will fully rotate around two axes over the course of a year (one normal to the ecliptic, the other inclined by 60 degrees), allowing its varying sky-dependent response function to provide additional source position information \citep{2017AmaroSeoane,Cornish2001}. 
The different arm orientations of the three legs of the constellation, and their variation due to the orbit, both provide information about the parameters of the emitting bodies. 
LISA's sweeping orbit and the longer time periods that its sources remain in-band will allow it to gather similar information to the suite of planned ground detectors with a single instrument \citep{2017AmaroSeoane}. Variations on the nominal LISA design are of interest both for followup mission concepts, as well as contemporaneous concepts like TianQin \citep{Luo2016} and Taiji \citep{Ruan2018}.

The deployment of space-based interferometric GW detectors requires reducing extraneous non-gravitational accelerations, referred to as acceleration noise, as much as possible, to best approximate a measurement with free-falling test masses. 
The effect of spurious random forces with respect to the spacecraft's inertial frame limit the instrument's sensitivity at low frequencies, whereas photon shot noise and other optical path noises limit the high-frequency sensitivity \citep{Armano2016}.The photon shot noise is the same phenomenon that limits ground-based detectors at high frequencies, and the acceleration noise as a limitation requires mitigation in order to reach the required sensitivity.

To filter out non-gravitational forces as much as possible, each LISA spacecraft uses two free-floating test masses as geodesic reference test particles. 
Each spacecraft is then forced to follow its test mass using micro-Newton thrusters. 
After applying these corrections to the spacecraft, the distance between test masses along each arm of the interferometer is calculated, with corrections for Doppler shifts and post-processing on Earth \citep{2017AmaroSeoane}. 
The data from the three arms (and six laser links) can be combined in a number of different ways to form sets of three different data channels, although a primary goal of most combinations is to cancel out the effect of the relative orbital motion of the satellites, a technique referred to as time-delay interferometry (TDI). 
In one such combination, the resulting set of observables is similar to three separate Michelson interferometers, each of which is comparable to a single ground-based detector.
For this work, we specifically calculate the SNR of the Michelson X channel of the XYZ TDI combination, but one can straightforwardly modify to calculate other TDI combinations \citep{Tinto2014,Smith2019}, although the network SNR should be the same regardless.

\subsection{LISA Power Spectral Density}{\label{subsec:LISAPSD}}
    The noise limitations on LISA in the high frequency regime are determined by laser interferometer noise (IFO), consisting of shot noise and any other optical path noises (e.g. the telescope, optical bench, phase measurement system, laser, clock and TDI processing), and is combined and modeled as
    \begin{equation}
        \label{eq:LISAIFOPSD}
        P_{\mathrm{IFO}}=\left(1.0 \times 10^{-12} \mathrm{m}\right)^{2}\left(1+\left(\frac{2\, \mathrm{mHz}}{f}\right)^{4}\right) \mathrm{Hz}^{-1},
    \end{equation}
    which is the proposed effective total displacement noise in one direction between two of the LISA spacecraft arms from the L3 mission proposal \citep{2017AmaroSeoane}. This noise source can be expressed more generally as
    \begin{equation}
        \label{eq:LISAIFOPSDModel}
        P_{\mathrm{IFO}}=\left(A_{\mathrm{IFO}}\right)^{2}\left(1+\left(\frac{f_{\mathrm{break}}}{f}\right)^{4}\right) \mathrm{Hz}^{-1},
    \end{equation}
    where $A_{\mathrm{IFO}}$ is the displacement amplitude of a single-link, and $f_{\mathrm{break}}$ is the frequency of the turnover for the high frequency noise.

    The low frequency regime is dominated by spacecraft acceleration noise \citep{Larson2000}.
    This is represented in the L3 proposal by
    \begin{equation}
        \label{eq:LISAaccPSD}
        P_{\mathrm{acc}}=\left(3 \times 10^{-15} \mathrm{m~s}^{-2}\right)^{2}\left(1+\left(\frac{0.4\, \mathrm{mHz}}{f}\right)^{2}\right)\left(1+\left(\frac{f}{8 \,\mathrm{mHz}}\right)^{4}\right) \mathrm{Hz}^{-1} ,
    \end{equation}
    which can be expressed more generally as
    \begin{equation}
        \label{eq:LISAaccPSDModel}
        P_{\mathrm{acc}}=\left(A_{\mathrm{acc}} \right)^{2}\left(1+\left(\frac{f_{\mathrm{break,low}}}{f}\right)^{2}\right)\left(1+\left(\frac{f}{f_{\mathrm{break,high}}}\right)^{4}\right) \mathrm{Hz}^{-1} .
    \end{equation}
    When looking at LISA's noise curve (i.e., figure \ref{fig:LISASensitivity}), the dependencies from equations \ref{eq:LISAIFOPSD} and \ref{eq:LISAaccPSD} determine the entire sensitivity curve \citep{2017AmaroSeoane}.

    If we assume each detector in the LISA configuration has the same noise spectral density, we can just add additional factors of each noise type.
    In a LISA-like interferometer there are four contributions to the interferometer noise and sixteen contributions to the acceleration noise. 
    Half of the acceleration noise contributions are scaled by $\cos^{2}(f/f_{*})$, where $f_{*} = c/(2 \pi L)$ is the round-trip light frequency (i.e., the transfer frequency), to account for combining the acceleration noise at two different times \citep{Cornish2002}.
    To combine the high and low frequency noise into strain, we must convert from displacement to strain by dividing by the square of the round-trip light travel distance, $2L$ \citep{Robson2019}.
    Furthermore, to obtain the effective position noise due to spurious accelerations, we must divide the total acceleration noise by the angular frequency of the GW, $2\pi f$, to the fourth power \citep{Cornish2002}.
    The total noise spectral density is then,
    \begin{equation}
        \label{eq:LISAPSD}
        P_{n}(f)=\frac{4P_{\mathrm{IFO}}}{(2L)^{2}}+8\left(1+\cos ^{2}\left(f / f_{*}\right)\right) \frac{P_{\mathrm{acc}}}{(2 \pi f)^{4} (2L)^{2}} .
    \end{equation}
    
\subsection{The Transfer Function}
\begin{figure}[!htbp]
        \includegraphics[width=\textwidth]{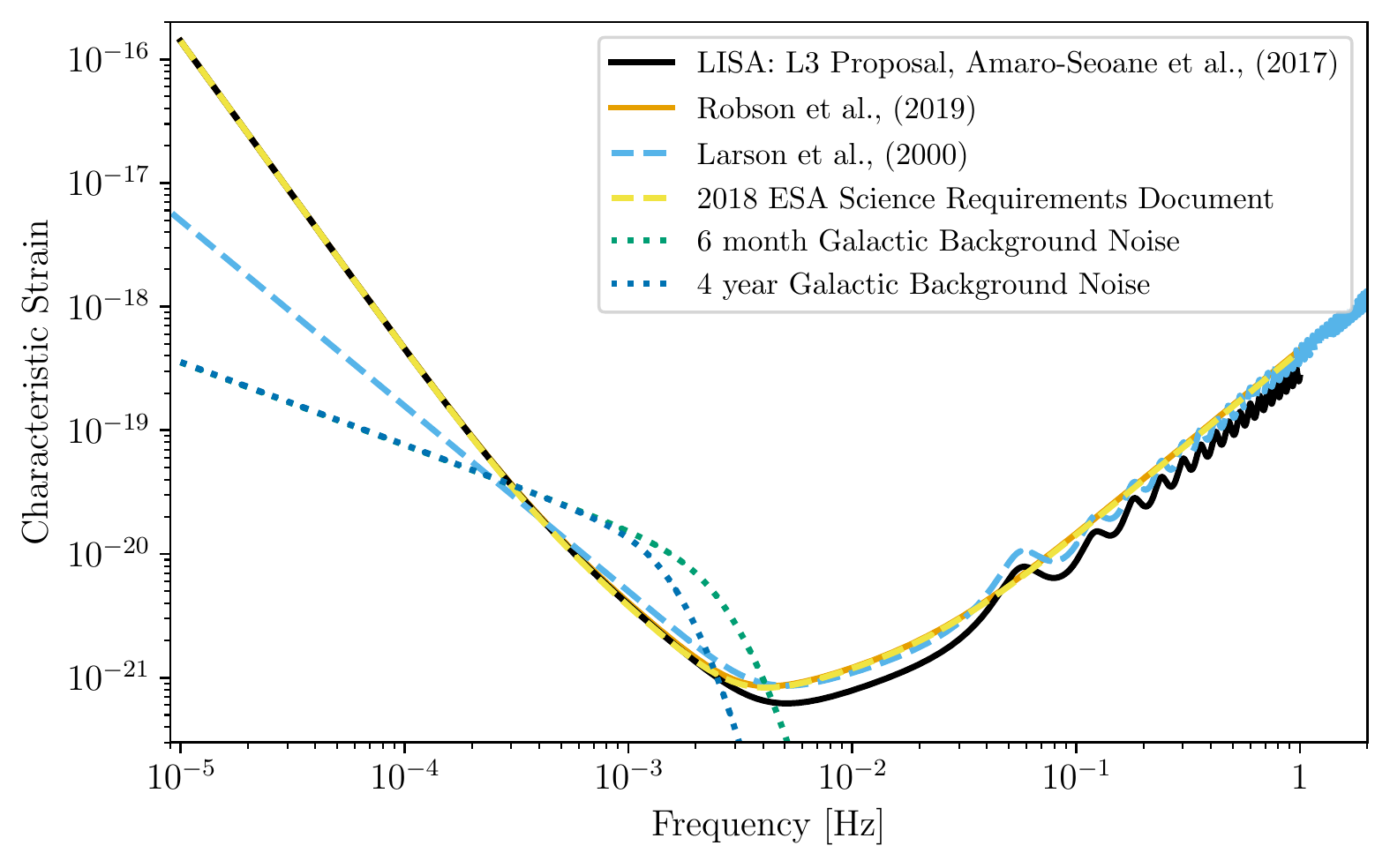}
    
        \caption{Sensitivities of various proposed LISA designs along with the galactic binary confusion noise. Differences between proposals and models made with \texttt{gwent} mostly change with respect to low-frequency spectral indices and high-frequency interferometer noise floors. The models for the 2018 ESA Science Requirements Document and \cite{Robson2019} use the approximation shown in equation \ref{eq:approxLISAResponse}. }
        \label{fig:LISASensitivity}
    \end{figure}

{\label{subsec:TransferFunc}}
    As mentioned in \S\ref{susubsec:Response}, instrument sensitivities depend on the source's properties.
    Additionally, the response functions of interferometers are dependent on the arm length of the detector in relation to the amplitude of the passing GW.
    This effect is encapsulated in the transfer function,
    \begin{equation}
        \label{eq:LISATransfer}
        \mathcal{T}(\mathbf{r} \cdot \widehat{\Omega}, f)=\operatorname{sinc}\left[\frac{f}{2 f_{*}}(1-\mathbf{r} \cdot \widehat{\Omega})\right] e^{i \frac{f}{2 f_{*}}}(1-\mathbf{r} \cdot \widehat{\Omega}) .
    \end{equation}
    where $\mathbf{r}$ is a unit vector pointing from the first to the second spacecraft \citep{Cornish2001}.
    For $f \ll f_{*}$, the transfer function approaches unity, while at frequencies higher than $f_{*}$, the sinc function causes periodic dips at $f=n/2L$ and is proportional to $1/f$ for $f \gg f_{*}$ \citep{Larson2000}.
    This can be seen in figure \ref{fig:LISASensitivity} for the models with numerically calculated transfer functions.
    
    The response function in equation \ref{eq:angavgstrain} is made up of the instrument antenna functions
    \begin{equation}
        F^{A}(\widehat{\Omega}, f)=\mathbf{D}(\widehat{\Omega}, f) : \mathbf{e}^{A}(\widehat{\Omega}) ,
    \end{equation}
    where the colon denotes the double contraction $\mathbf{a} : \mathbf{b}=a_{i j} b^{i j}$.
    The antenna pattern is made up of the detector tensor
    \begin{equation}
        \label{eq:LISAdetectortensor}
        \mathbf{D}(\widehat{\Omega}, f)=\frac{1}{2}(\mathbf{r} \otimes \mathbf{r}) \mathcal{T}(\mathbf{r} \cdot \widehat{\Omega}, f) ,
    \end{equation}
    and the basis tensors of the GW polarizations $\mathbf{e}^{A}(\widehat{\Omega}$) are given by equation \ref{eq:BasisTensors}.

    Because of the complexity of the functions in $\mathcal{R}(f)$ from equation \ref{eq:angavgstrain}, one needs to numerically integrate to get the full LISA response function.
    We use the method described in \cite{Larson2000} to produce the LISA response curves in this paper and provide the result in a text file inside of \texttt{gwent}.
    
    Alternatively, one can use an analytic fit for the response function
    \begin{equation}
        \label{eq:approxLISAResponse}
        \mathcal{R}(f)=\frac{2 \sin^{2}\beta}{5} \frac{1}{\left(1+0.6\left(f / f_{*}\right)^{2}\right)} ,
    \end{equation}
    found in \cite{Robson2019}, where the $2\, \sin^{2}\beta/5$ prefactor ensures that $\mathcal{R}(f)$ asymptotes to the appropriate low frequency limit, where $\beta$ is the opening angle between two detectors \citep{Cornish2001}.
    The normalization in equation \ref{eq:approxLISAResponse} is a factor of 2 larger than what is given in some of the literature, since it accounts for a sum over the two independent low-frequency data channels \citep{Robson2019}).
    
\subsection{LISA Sensitivity Curve}
    We find LISA's effective noise PSD,
    \begin{align}
        \label{eq:LISAENPSD}
        S_{n}(f) &= \frac{P_{n}(f)}{\mathcal{R}(f)} \\
        &= \frac{10}{3\mathcal{R}(f)}\left(\frac{P_{\mathrm{IFO}}}{L^{2}}+2\left(1+\cos^{2}\left(f/f_{*}\right)\right) \frac{P_{\mathrm{acc}}}{(2\pi f)^{4}L^{2}}\right)
    \end{align}
    by using LISA's PSD from equation \ref{eq:LISAPSD} and its response function from either the stored result of the numerical integration in \citep{Larson2000} or from equation \ref{eq:approxLISAResponse}.
   
\subsubsection{Galactic Binary Background}
         Current population modeling of interacting white dwarf binaries estimate around 26 million binaries emit GWs in the mHz band \citep{Cornish2017}.
        Most of these binaries are unresolved, so that for the purpose of observing BHBs, they constitute an additional noise source and their power should be added to the instrument noise power, in direct analogy to the role of the GWB when trying to observe CWs with PTAs.
        To represent this confusion noise, we use a modified version of the smooth fit to the co-added signals from the simulated Galactic binary population in \cite{Cornish2017}.
        We determined that most of the fit parameters employed in \cite{Cornish2017} were poorly constrained, even with the correction to the fit in \cite{Schmitz2020}.
        Since the parameters had only a very weak effect on the signal; this led to the highly non-monotonic behavior of three of the fit parameters when varying the observation time. 
        Therefore, we digitized the curves in Figure 2 of \cite{Cornish2017}, and fit each of them to the simpler function
        \begin{equation}
            \label{eq:WDback}
            S_{c}(f)=A f^{-7 / 3} \left[1+\tanh \left(\gamma\left(f_{k}-f\right)\right)\right] \mathrm{Hz}^{-1}\,.
        \end{equation}
        We then fit $A$, $\gamma$, and $f_{k}$ as power laws of $T_{\rm obs}$, finding $A=1.4 \times 10^{-44}$, $\gamma=1100\,\left(T_{\rm obs}/1\,{\rm yr}\right)^{3/10}$, and $f_k = 0.0016\,\left(T_{\rm obs}/1\,{\rm yr}\right)^{-2/9}$ Hz. 
        We note that this fitting function is physically motivated in the asymptotic limits; it is simply the $f^{-7/3}$ spectrum expected for a fixed population of binaries evolving slowly due to gravitational radiation at low frequencies, and is exponentially suppressed as we move to frequencies $f \gg f_k$ since an increasing fraction of binaries in each frequency bin becomes individually resolvable \cite{Barack2004}. 
        The $\tanh$ function provides a simple transition between these two limits.
 
        We show the resulting characteristic strain in figure \ref{fig:LISASensitivity}.
        Summing the PSD of the Galactic WD binaries and that of the instrumental noise allows us to represent the total sensitivity. 
        We note that in the LISA context, the response function is negligible at the low frequencies where Galactic binaries dominate, but for designs with longer arm lengths, the effects may occur in the same band, so it is important to apply the response function to both the galactic foreground and the instrument noise consistently (i.e., $S_{n}(f) = \left(P_{n}(f) + S_{c}(f)\right)/\mathcal{R}(f)$).
    
\subsection{Sensitivity Changes to Various Designs}
    Because of the versatility of our sensitivity model defined in equation \ref{eq:LISAPSD}, using \texttt{gwent}'s SNR calculation capabilities, we can explore the wide parameter space of each instrument variable.
    The differences among the models in figure \ref{fig:LISASensitivity} are due to variations in the low-frequency spectral index, the interferometer noise floor, and exact versus approximate transfer functions.
    
    \begin{figure}[!htbp]
        \includegraphics[width=\textwidth]{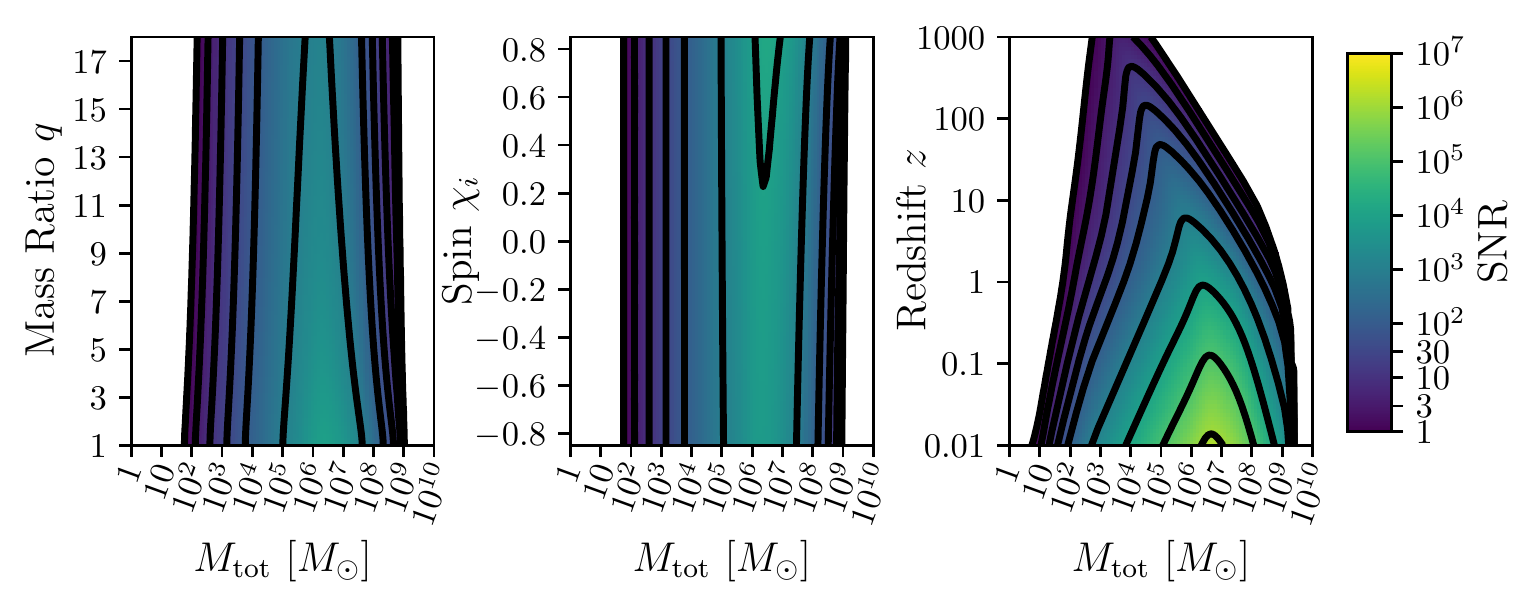}
    
        \caption{L3 ESA LISA model SNRs for changes to a fiducial, non-spinning, equal-mass BHB at a redshift of one ($\chi_{i}=0, z=1, q=1$). \textbf{Left:} mass ratio vs total source mass, \textbf{Middle:} spins of both BHs $\chi_{i}=\chi_{1}=\chi_{2}$ vs total source mass, \textbf{Right:} redshift vs total source mass.
        Solid black contours correspond to the levels marked on the colorbar.}
        \label{fig:LISA_M_vs_source_param}
    \end{figure}

    In figure \ref{fig:LISA_M_vs_source_param} we show the sensitivity to a fiducial non-spinning, equal mass example BHB at a redshift of one ($\chi_{i}=0, z=1, q=1$) in the LISA band, to changes in those source parameters versus total mass.
    We find that for mass ratio, the L3-proposed LISA configuration covers a consistent parameter space in mass no matter the ratio.
    There is an increase in signal strength for more equal mass BHBs.
    Similarly, the detectability of this example BHB source spin maintains the same total detectable space in source mass, but has a preference in strength to highly aligned spin.
    We note that the largest SNR with respect to the marginal effect of changing the spin of each black-hole in the binary happens when the merger occurs over the most sensitive frequencies.
    For LISA's response to BHBs at different redshifts, we note a similar trend; the region of highest SNR is dominated by sources that merge over the detector's most sensitive frequencies. 
    Much like PTAs, LISA has the potential to see particular BHBs coalescing at the earliest times in the Universe, in the unlikely case that they exist.
    
\begin{figure}[!htbp]
        \includegraphics[width=\textwidth]{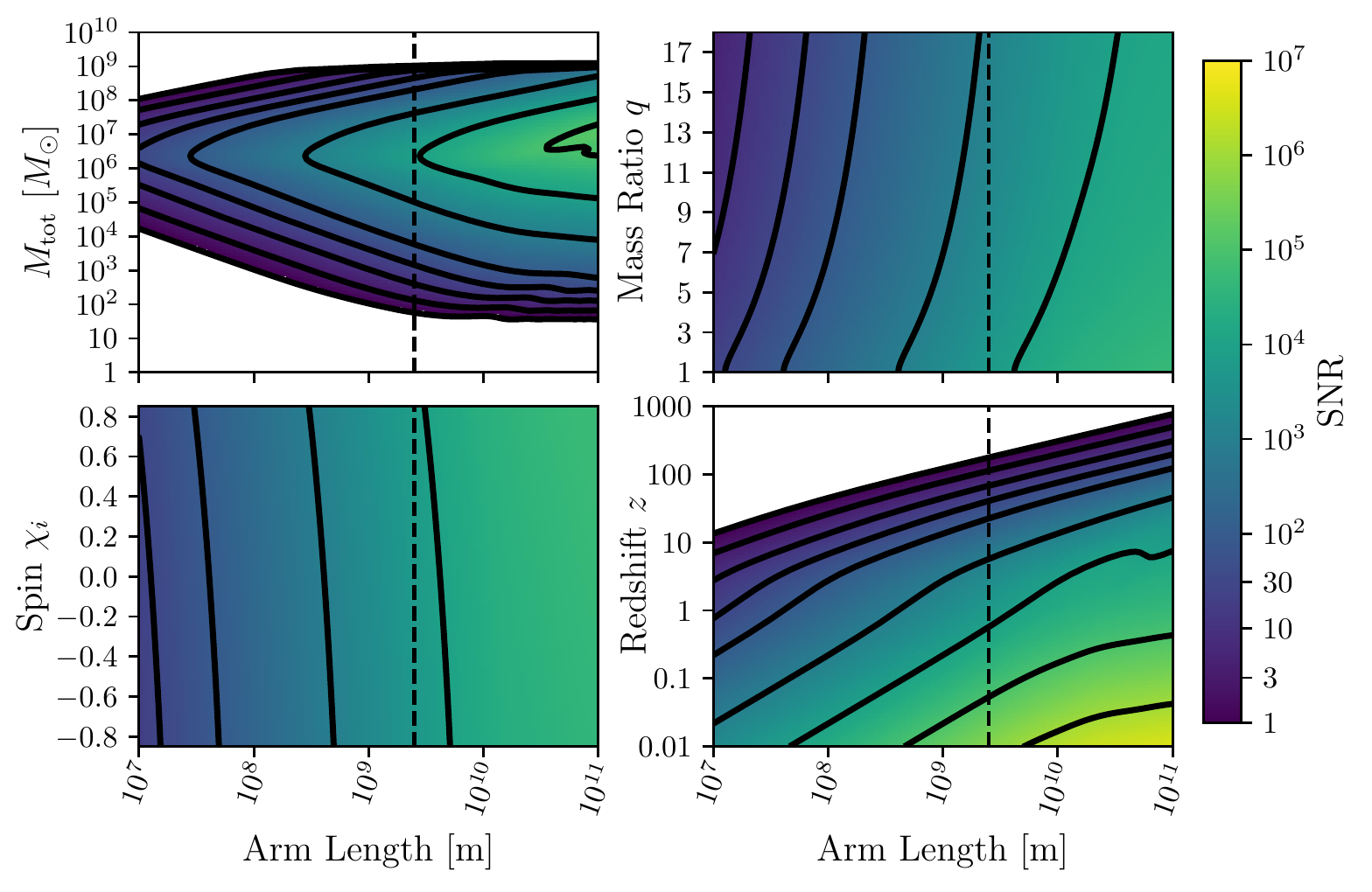}
    
        \caption{SNRs for LISA arm lengths compared to changes to a fiducial BHB ($M_{\mathrm{tot}}=10^{6} \mathrm{M}_{\odot}, \chi_{i}=0, z=1, q=1$). Dashed lines represent the proposed value for the arm length. Solid black contours correspond to the levels marked on the colorbar. \textbf{Upper Left:} total source mass vs arm length, \textbf{Upper Right:} mass ratio vs arm length \textbf{Lower Left:} spins of both BHs $\chi_{i}=\chi_{1}=\chi_{2}$ vs arm length \textbf{Lower Right:} redshift vs arm length. }
        \label{fig:LISAparamComp}
    \end{figure}

    In figure \ref{fig:LISAparamComp}, we look at the same fiducial source BHB as figure \ref{fig:LISA_M_vs_source_param} ($M_{\mathrm{tot}}=10^{6}~ \mathrm{M}_{\odot}, \chi_{i}=0, z=1, q=1$) as a function of both the detector's arm length and various source parameters.
    It is evident that the arm length greatly impacts the sensitivity of the detector.
    In general, a longer arm length increases the sensitivity across the entire source parameter space.
    
    Increasing the arm length can expand the detectable total BHB mass space to both higher and lower masses until around the arm length planned for LISA.
    Above this arm length, there appears to be a point of diminishing returns in terms of the observable mass range, although the sensitivity to a given mass continues to increase.
    As expected, the SNR increases with increasing spin and decreasing (i.e., closer to unity in our convention) mass ratio, with mass ratio being a stronger factor.
    As the arm length increases, LISA steadily gains sensitivity out to larger redshifts.
    While this analysis would suggest that a larger constellation is always better, we note that this ignores the increased cost of reaching those orbits, and the potential lack of sufficiently stable orbits until one reaches AU-scale constellations (see e.g.~\cite{Folkner2011}).
    
    \begin{figure}[!htbp]
        \includegraphics[width=\textwidth]{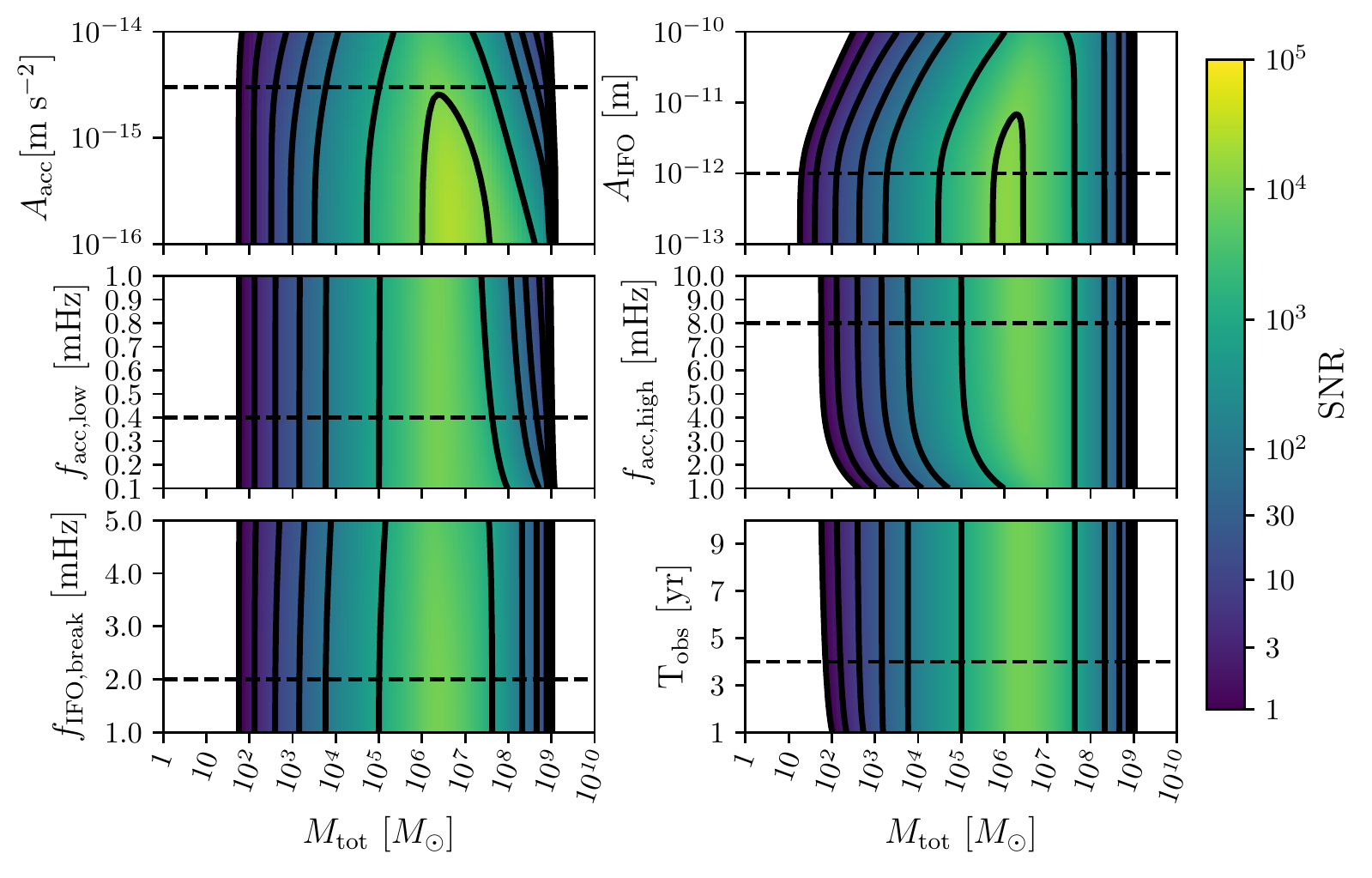}
    
        \caption{SNRs for a fiducial, non-spinning, equal-mass BHB at a redshift of one ($\chi_{i}=0, z=1, q=1$) compared to changes to the noise parameters for the L3 ESA LISA model. \textbf{Top Left:} acceleration noise amplitude, $A_{\mathrm{acc}}$ vs total source mass. Dashed lines represent the fiducial values for each parameter. Solid black contours correspond to the levels marked on the colorbar. \textbf{Top Right:} optical metrology noise amplitude, $A_{\mathrm{IFO}}$ vs total source mass. \textbf{Middle Left:} low frequency acceleration noise break frequency, $f_{\mathrm{acc, low}}$ vs total source mass. \textbf{Middle Right:} high frequency acceleration noise break frequency, $f_{\mathrm{acc, high}}$ vs total source mass. \textbf{Bottom Left:} optical metrology noise break frequency, $f_{\mathrm{IFO, break}}$ vs total source mass. \textbf{Bottom Right:} observation time length $T_{\mathrm{obs}}$ vs total source mass.}
        \label{fig:LISA_M_vs_inst_param}
    \end{figure}

    In addition, sensitivity does not always correlate with parameter measurement accuracy; the larger orbits would modulate a signal more slowly, but would also have a nontrivial frequency dependent response over more of their sensitive band, and these features would compete with one another in terms of affecting measurement accuracy. We therefore emphasize that our calculations can only address the question of observability of given sources, assuming such design parameters can be realized in practice, and does not necessarily reflect what design would yield the most accurate measurements.
    
    In figure \ref{fig:LISA_M_vs_inst_param}, we look at the dependence of the sensitivity to a non-spinning, equal-mass BHB at $z=1$, to variations of the other modeled LISA parameters (using the noise values for the L3 ESA LISA proposal given in equations \ref{eq:LISAaccPSD} and \ref{eq:LISAIFOPSD}).
    In most cases, the detectable total mass parameter space stays constant.
    In fact, reasonably reducing the noise levels often only marginally increases the SNR.
    This kind of examination shows that the most impactful noise parameter for LISA is its arm length, as seen in figure \ref{fig:LISAparamComp}.

\section{Ground Based Detector Sensitivity}
\label{sec:LIGOsec}
In general, the current generation of ground-based GW detectors have a frequency range of $\sim$10 Hz to a few kHz. 
This high frequency band primarily contains compact binaries (some pairing of stellar-mass black-holes and neutron stars) and supernovae sources. Being attached to the Earth, ground-based detectors have significant sources of environmental noise.
The lower frequency limit has a steep cutoff due to seismic noise (i.e., vibrations due to the ground motion that cannot be eliminated by the suspension system); design choices like improved seismic isolation and longer arm lengths can somewhat mitigate this noise, but it is a significant challenge to push below the limit due to the Newtonian background (the time-varying mass distribution near the test mass) of a few Hz for terrestrial detectors \citep{Pepper2017,Saulson1994}.
For the low to mid-range frequencies, there are a combination of noise levels, but the sensitivity is primarily limited by gravity gradient noise and quantum optical noise (photon radiation pressure noise for lower frequencies) \citep{Hild2011}.
At the higher frequencies, the noise is dominated by photon shot noise, which, along with other potential optical path noises, determine the sensitivity for both ground-based interferometers like LIGO/Virgo \citep{Abbott2016} and KAGRA \citep{Somiya2012}, and space-based interferometers like LISA \citep{2017AmaroSeoane} at the high end of their respective bands.   

\begin{figure}[!htbp]
    \includegraphics[width=\textwidth]{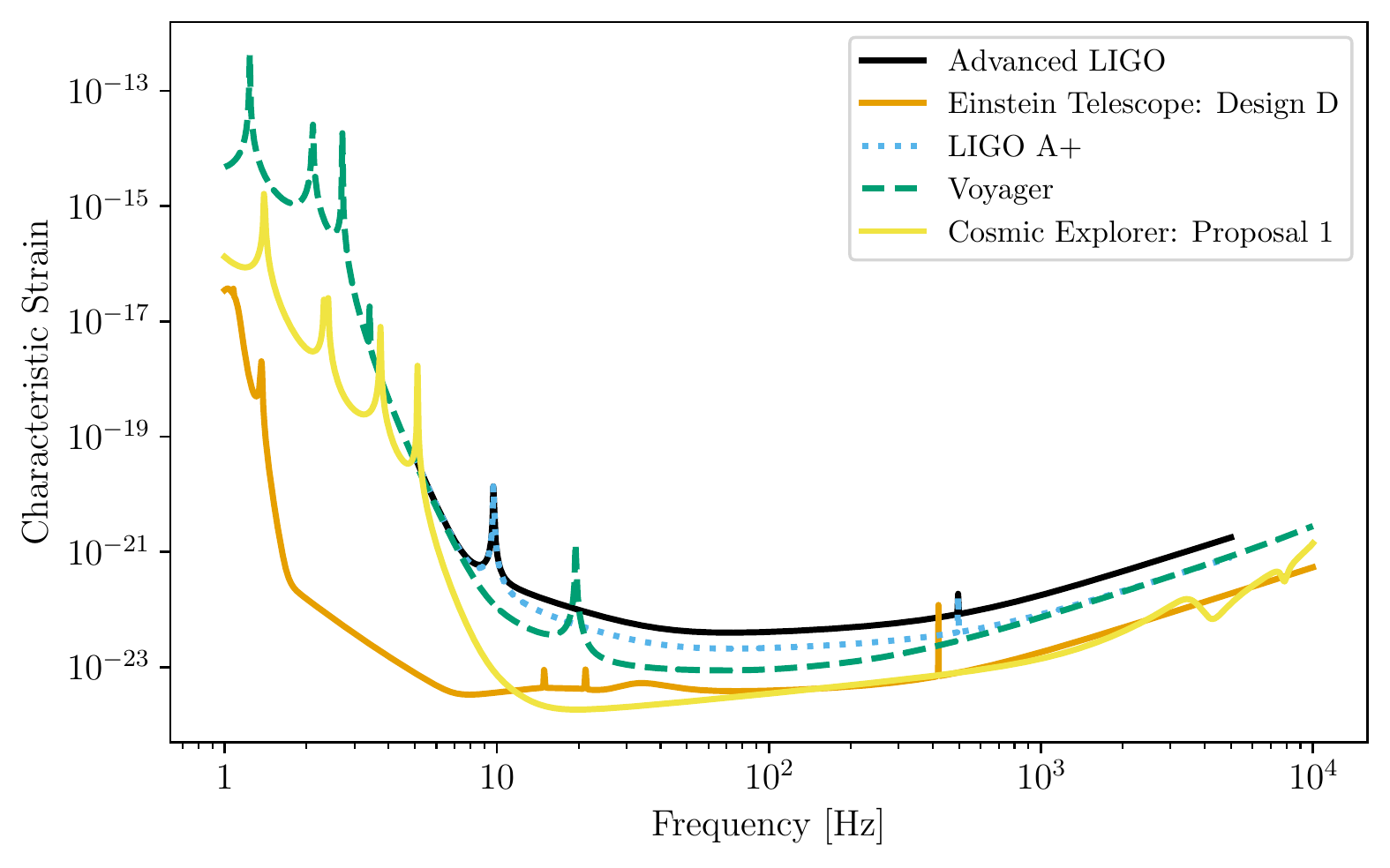}

    \caption{Sensitivities of various ground detector designs. 
    }
    \label{fig:GroundSensitivity}
\end{figure}

\subsection{Sensitivity Measurements}
    In principle, we can treat ground-based interferometer sensitivities much like space-based ones.
    Ground-based detectors also have various noise sources like thermal, seismic, quantum, etc. whose power can be summed to generate a sensitivity curve \citep{Hild2011}.
    Ground-based instruments also need to take into account the arm length penalty expressed in \S\ref{subsec:TransferFunc} if their arms are long enough.
    The transfer frequency for aLIGO is $f_{*}\approx10^{4}$ Hz, thus we can safely assume current ground-based detectors operate in the low frequency limit of a flat response function.
    Future detectors that are more sensitive at higher frequencies and/or have longer arms will need to include the full response function, as seen in the Cosmic Explorer sensitivity curve in figure \ref{fig:GroundSensitivity}.
    
    \begin{figure}[!htbp]
        \includegraphics[width=\textwidth]{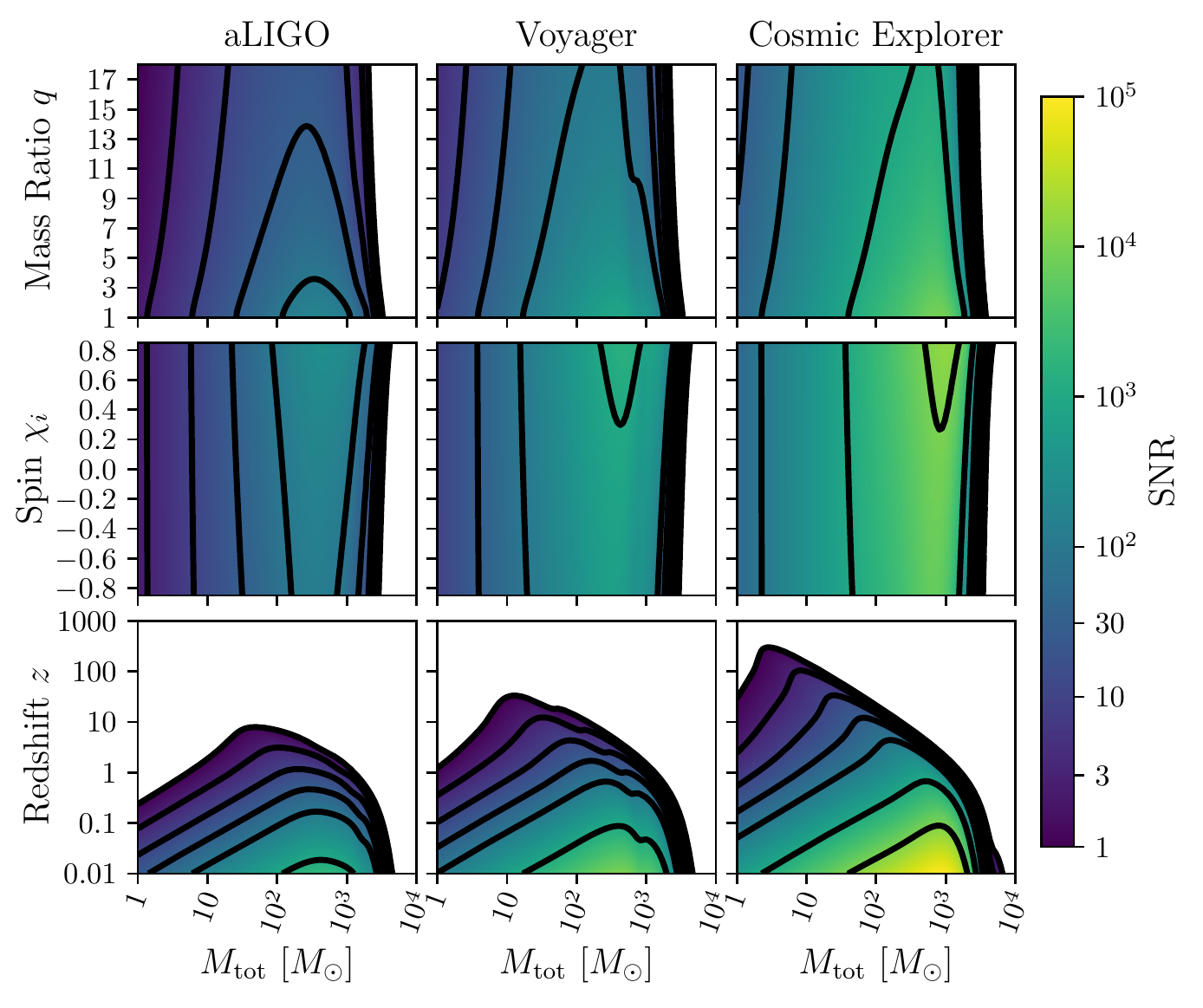}
    
        \caption{SNRs for changes to a fiducial, non-spinning, equal-mass BHB at a redshift of one ($\chi_{i}=0, z=1, q=1$) with respect to ground-based detectors: aLIGO (Left Column), Voyager (Middle Column), and Cosmic Explorer (Right Column). Solid black contours correspond to the levels marked on the colorbar. \textbf{Top Row:} mass ratio vs total source mass, \textbf{Middle Row:} spins of both BHs $\chi_{i}=\chi_{1}=\chi_{2}$ vs total source mass, \textbf{Bottom Row:} redshift vs total source mass.
        }
        \label{fig:GroundparamComp}
    \end{figure}
    
    Because of the extensive studies of ground-based sensitivities, there exist tools to model both the micro and macro noise effects in these instruments.
    Thus one can better characterize noise inside the instrument and better model current and future detectors \citep{Martynov2016,Miller2015}.
    For all of our SNR calculations and the ground-based sensitivity curves, we use \texttt{pygwinc}\footnote{\href{https://git.ligo.org/gwinc/pygwinc/tree/master}{https://git.ligo.org/gwinc/pygwinc/tree/master}} with the lone exception of the sensitivity estimate for ET, for which we use publicly available data\footnote{\href{http://www.et-gw.eu/index.php/etsensitivities}{http://www.et-gw.eu/index.php/etsensitivities}}.
    
    \texttt{pygwinc} is an analytic sensitivity curve generator for current and future ground-based GW detectors.
    It combines both public and collaboration-internal noise estimates to provide designs with the most up-to-date information available.
    For these reasons, we implement \texttt{pygwinc} as the primary tool for generating ground-based sensitivity curves.
    The generated sensitivity curves are integrated into \texttt{gwent}'s framework to provide full model adaptability to each variable parameter and ground-based detector design.
    
    In figure \ref{fig:GroundSensitivity} we display the default configurations for aLIGO, LIGO A$+$, Voyager, and Cosmic Explorer using this formalism, along with 
    the sensitivity for ET taken from publicly available data.

\subsection{Sensitivity Changes to Various Designs}
    We examine the detectable GW source parameter space with this formalism using the fiducial detector designs.
    In figure \ref{fig:GroundparamComp} we show the results of varying the mass ratio, spin, and redshift versus the total mass of a BHB in the ground detector mass regime.
    It should be noted that all detectable BHBs in the explored parameter space will be evolving in frequency.
    We find that there is only a moderate bias in SNR strength towards an aligned, highly-spinning BHB in both the aLIGO and ET models.
    Both detectors exhibit the consistent behavior we have seen for all GW detectors in this work of higher SNRs for closer to equal-mass mergers.
    
    \begin{figure}[!htbp]
        \includegraphics[width=\textwidth]{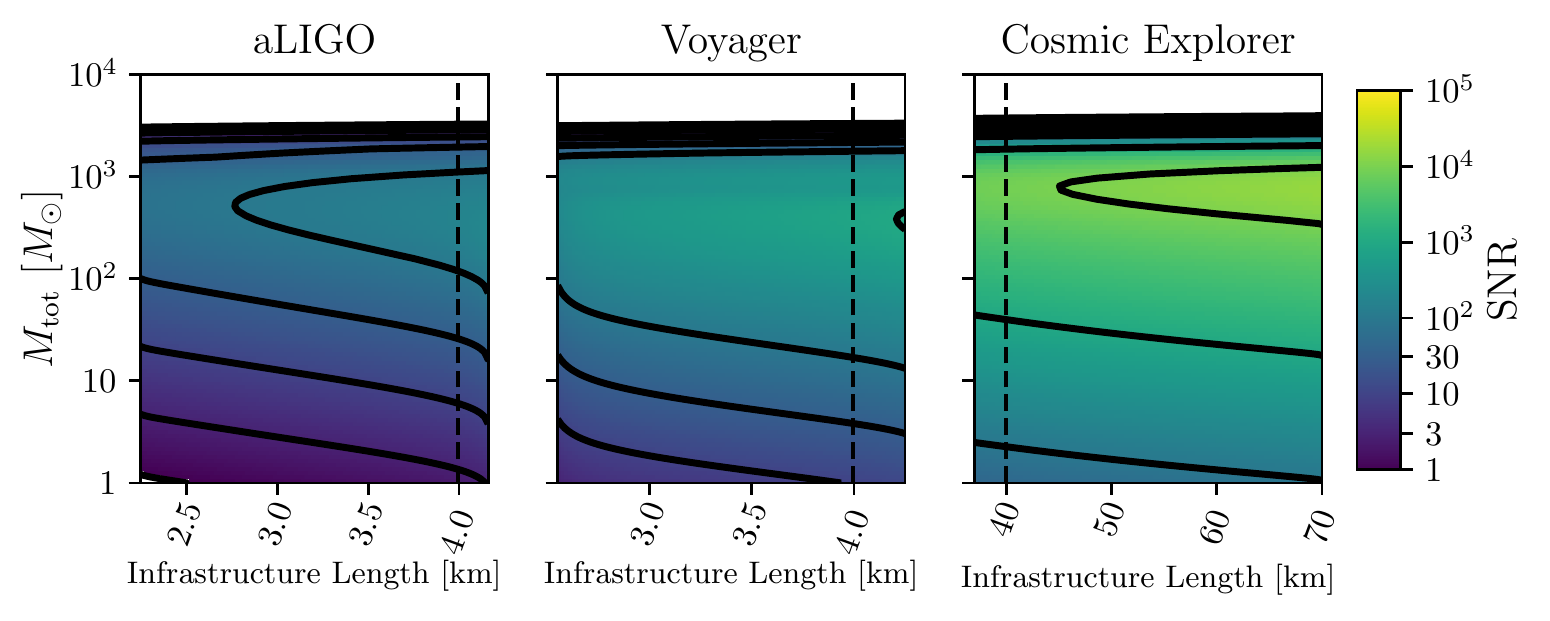}
        \caption{SNRs for the response of variations in arm length to a fiducial, non-spinning, equal-mass BHB at a redshift of one ($\chi_{i}=0, z=1, q=1$) for aLIGO (\textbf{Left Column}), Voyager (\textbf{Center Column}), and CE (\textbf{Right Column}). Black dashed lines represent the fiducial arm lengths in \texttt{pygwinc}. Solid black contours correspond to the levels marked on the colorbar.
        }
        \label{fig:Ground_M_vs_L}
    \end{figure}
    
    By extending the arms or making other modifications to the nominal design of current or proposed detectors, we can extend the sensitivity to lower frequencies and/or improve the sensitivity across the band, thereby increasing the SNRs, horizon distances, and sensitive mass ranges.
    In some cases, the change in sensitivity can be one or more orders of magnitude relative to aLIGO.
    
\begin{figure}[!htbp]
        \includegraphics[width=\textwidth]{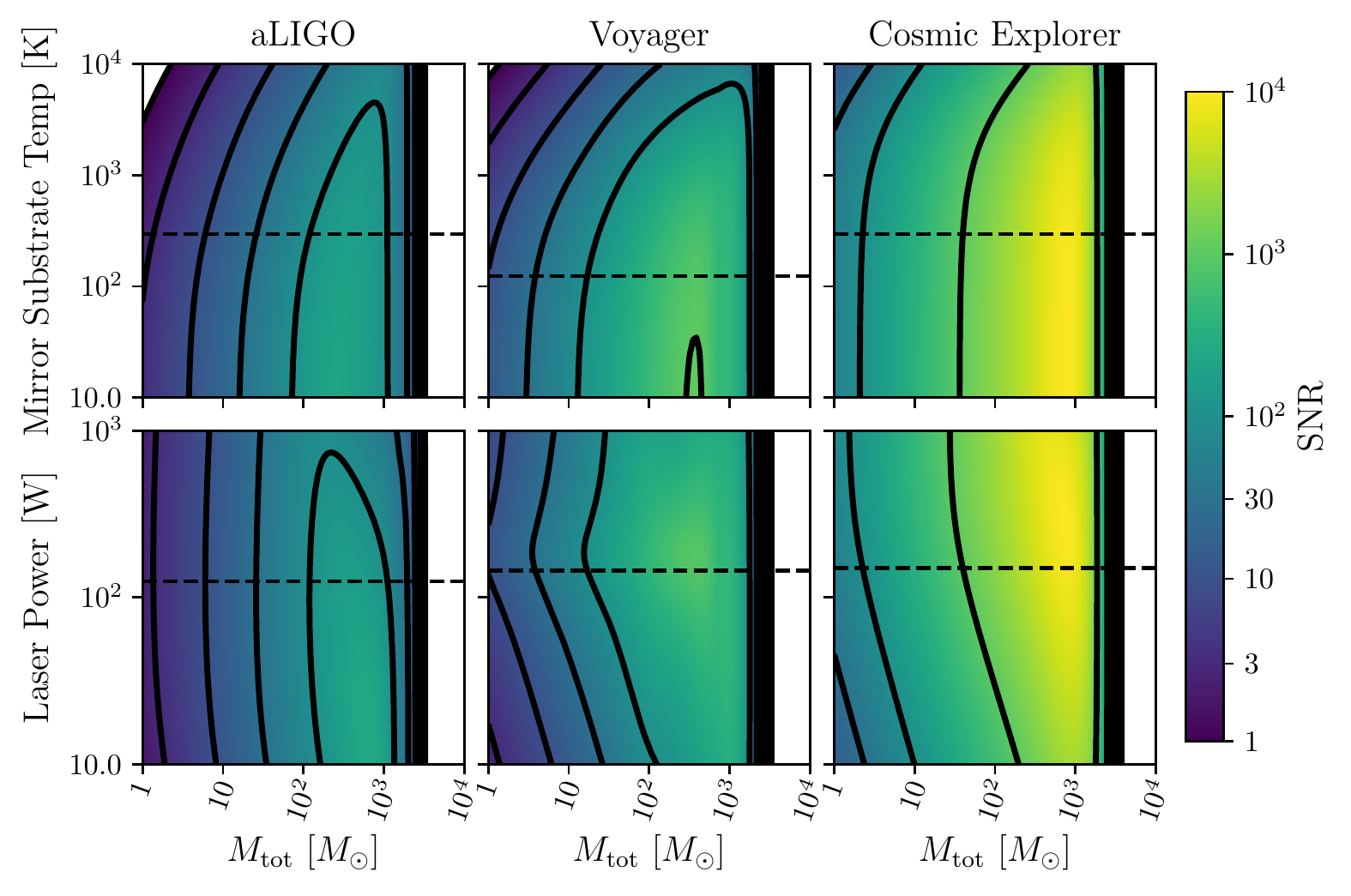}
        \caption{SNRs for the response of variations in instrument parameters to a fiducial, non-spinning, equal-mass BHB at a redshift of one ($\chi_{i}=0, z=1, q=1$) for aLIGO (\textbf{Left Column}), Voyager (\textbf{Center Column}), and CE (\textbf{Right Column}). 
        Black dashed lines represent the fiducial noise parameters in \texttt{pygwinc} for each detector. 
        Solid black contours correspond to the levels marked on the colorbar.
        \textbf{Top Row:} mirror substrate temperature vs total source mass. \textbf{Bottom Row:} laser power vs total source mass.}
        \label{fig:Ground_inst_param_v_M}
    \end{figure}

    We explore varying several of these instrument design parameters around the fiducial instrument values for aLIGO, Voyager, and CE in figures \ref{fig:Ground_M_vs_L} and \ref{fig:Ground_inst_param_v_M}.
    We find that, for the particular source of a non-spinning, equal-mass BHB at a redshift of $z=0.1$, the SNR for this example source increases with the arm length for all three detector designs, shown in figure \ref{fig:Ground_M_vs_L}.
    When increasing the arm length (and some other parameters) much outside of the initial design, one needs to simultaneously change other parameters in order to stay in the allowed region of detector parameter space as enforced by \texttt{pygwinc}.
    In this work, we do not simultaneously vary parameters and merely limit the range of variability to the allowed region around the proposed instrument's design value.
    
    Furthermore, for this particular BHB, there is a significant decrease in sensitivity to total mass when increasing the temperature of the mirror substrate.
    This trend, along with the sensitivity gains leveling out at temperatures below the proposed design levels is common to all of the simulations seen in figure \ref{fig:Ground_inst_param_v_M}.
    This tendency to reach a point of diminishing returns in sensitivity is common to many parameters because as the individual component's noise floor lowers (in this case by decreasing the substrate temperature), the noise component contributes less to the overall noise as other non-varying sources come to the foreground.
    
    In terms of laser power, we see in figure \ref{fig:Ground_inst_param_v_M} that aLIGO's sensitivity depends only weakly on the laser power near its design value. In contrast, there appears to be an optimal laser power that maximizes the SNR for particular mass ranges of the example BHB source for Voyager and CE.
    The optimal value of $\sim200$ W for Voyager occurs near its actual planned operating value.
    For CE, increasing the laser power to the upper bound of the planned operating range yields a steady increase in SNR with increasing laser power for stellar-mass binaries, whereas for a narrow range of intermediate-mass BHBs near $10^3\,M_{\odot}$, a lower laser power of $\sim300$ W maximizes the SNR. 
    Given the scientific value of such sources as a potential evolutionary missing link between stellar BHBs and SMBHBs, there may be significant benefit in spending some time operating CE closer to this optimal value.
    Additional investigations for other sources and detector arrangements are left for future work. 
    
    As demonstrated in figure \ref{fig:Ground_inst_param_v_M}, investigating the interplay of instrument parameters and BHB source parameters with this formalism can reveal optimal design values. 
    This again highlights the usefulness of the unified formalism implemented here.

\section{Discussion}
\label{sec:Discussion}
\subsection{Application of Sensitivities}
    We have presented the tool, \texttt{gwent}, the first open source software for constructing sensitivity curves and performing SNR calculations for current and future GW detectors across the entire GW spectrum of coalescing BHBs.
    Each detector model contains a sufficiently detailed parameter space to adequately describe the characteristics of a wide variety of detector designs.
    We present our models for interferometer and PTA GW detectors in the same context to facilitate a fair comparison of capabilities.
    Additionally, we use a versatile source model to extend the capabilities of our tool to variations in mass, mass ratio, spins, orientation, and redshift.
    
    By providing a robust tool for both GW detectors and sources, we are able to clearly show the synergy between PTA, space-based, and ground-based detectors across the relevant parameter space and over the entire GW frequency band.

    While examples of similar methods and tools to this study exist (e.g., the work of \cite{MooreCole2015}\footnote{\href{https://gwplotter.com/}{https://gwplotter.com/}}, \cite{Katz2019}\footnote{\href{https://github.com/mikekatz04/BOWIE}{BOWIE: Binary Observability With Illustrative Exploration}}, and \cite{Smith2019}\footnote{\href{https://doi.org/10.5281/zenodo.3341817}{https://doi.org/10.5281/zenodo.3341817}}), we believe our treatment is more versatile and extensible as \texttt{gwent} is continually adaptable and not limited to a particular instrument.

\subsection{Multiband observations of BHBs}
    Studying the millihertz frequency band with LISA allows us to connect the mostly monochromatic SMBHBs in the PTA band to their merging counterparts, and to study the lower mass mergers at larger redshifts that built up the more massive PTA sources at lower redshifts.
    Furthermore, LISA is sensitive to the early inspiral of stellar mass BHBs that merge within the ground-based band, providing vital information about their initial evolution and properties; see figure \ref{fig:fullSNR} for the overlapping parameter space between instruments.
    
    Including realistic PTA curves using the tools of \cite{Hazboun2019} alongside other GW detectors allows for a fair comparison of detectors across the GW spectrum.
    This can elucidate the ability of PTAs to detect individually resolvable GW sources in tandem with space-based detectors, albeit at different times in the BHB's evolution.
    Furthermore, the modeling ability presented here provides a robust framework for assessing future detector concepts.
    
\subsection{Waterfall Plot Estimates}
    The framework we have implemented here allows us to calculate realistic SNRs using multi-parameter GW source models alongside adaptable GW detector designs.
    Because of the adaptability of both the detector and source models, we can explore changes in detector sensitivity to various sources, as a function of either source or detector parameters.
    This allows us to assess the optimal instrument parameter choices as a function of either the source population or the detector design.
    
    By combining tools like ours with population synthesis models, one can employ a robust detection probability analysis and source parameter estimation for various GW detectors \citep{Katz2020}.
    One can supplement and expand existing methods of PTA detection probability estimation using similar methods \citep{Kelley2018}. 
        
    The adaptability of each of these GW detector models allows us to investigate optimal configurations to maximize sensitivity to particular sources.
    We plan to investigate these various configurations in future papers.
    By properly modeling BHB sources and GW detectors across multiple bands and detector classes, we can more accurately assess the true prospects for multi-band GW astronomy.

\ack
We thank Maura McLaughlin and Jeff Hazboun for helpful feedback on the manuscript and insightful discussion.
This work was supported by NSF PFC Grant PHY-1430284 and NSF CAREER Grant PHY-1945130.

\vspace{5mm}
{\large \it Software:} 
Astropy \citep{astropy2013,astropy2018}, Hasasia \citep{Hazboun2019JOSS}, Matplotlib \citep{Matplotlib2007}, NumPy \citep{Numpy2006,Numpy2011}, Python \citep{Python2007,Python2011}, SciPy \citep{Scipy2020}.

\bibliographystyle{iopart-num}
\newcommand{\aap}{\textit{Astron. Astrophys.}}               
\newcommand{\aass}{\textit{Astron. Astrophys. Supp. Ser.}}       
\newcommand{\aipcp}{\textit{AIP Conf. Proc.}}                     
\newcommand{\aj}{\textit{Astron. J.}}                          
\newcommand{\ajp}{\textit{American Journal of Physics}}         
\newcommand{\apj}{\textit{Astrophys. J.}}                       
\newcommand{\apjs}{\textit{Astrophys. J. Supp. Ser.}}            
\newcommand{\apjl}{\textit{Astrophys.~J.~Let.}}               
\newcommand{\am}{\textit{Ann. Math.}}                          
\newcommand{\ap}{\textit{Ann. Phys.}}                          
\newcommand{\anyas}{\textit{Ann. N. Y. Acad. Sciences}}           
\newcommand{\arnps}{\textit{Annu. Rev. Nucl. Part. Sci.}}         
\newcommand{\araa}{\textit{Annu. Rev. Astron. Astrophys.}}       
\newcommand{\baas}{\textit{Bull. Am. Astron. Soc.}}       
\newcommand{\cqg}{\textit{Class. Quantum Grav.}}                
\newcommand{\cmp}{\textit{Commun. Math. Phys.}}                 
\newcommand{\cpc}{\textit{Computer Physics Communications}}     %
\newcommand{\CUP}{\textit{Cambridge University Press}}          %
\newcommand{\grg}{\textit{Gen. Rel. Grav.}}                     
\newcommand{\ijmpa}{\textit{Int. J. Mod. Phys. A}}                
\newcommand{\ijmpd}{\textit{Int. J. Mod. Phys. D}}                
\newcommand{\jap}{\textit{J. App. Phys.}}                       
\newcommand{\jcap}{\textit{JCAP}}                                
\newcommand{\jcp}{\textit{J. Comp. Phys.}}                      
\newcommand{\jpcs}{\textit{J. Phys. Conf. Ser.}}                 
\newcommand{\jmp}{\textit{J. Math. Phys.}}                      
\newcommand{\lnp}{\textit{Lecture Notes in Physics}}            %
\newcommand{\lrr}{\textit{Living Rev. Relativity}}              
\newcommand{\mpla}{\textit{Mod. Phys. Lett. A}}                  
\newcommand{\mnras}{\textit{Mon. Not. R. Astron. Soc.}}           
\newcommand{\njp}{\textit{New Jour. Phys.}}                     
\newcommand{\OUP}{\textit{Oxford University Press}}             %
\newcommand{\pla}{\textit{Phys. Lett. A}}                       
\newcommand{\plb}{\textit{Phys. Lett. B}}                       
\newcommand{\prpt}{\textit{Phys. Rept.}}                         %
\newcommand{\pr}{\textit{Phys. Rev.}}                          
\newcommand{\pra}{\textit{Phys. Rev. A}}                        
\newcommand{\prd}{\textit{Phys. Rev. D}}                        
\newcommand{\prl}{\textit{Phys. Rev. Lett.}}                    
\newcommand{\prsla}{\textit{Proc. R. Soc. London Ser. A}}         
\newcommand{\ptp}{\textit{Prog. Theor. Phys.}}                  
\newcommand{\ptps}{\textit{Prog. Theor. Phys. Suppl.}}           
\newcommand{\ptrsl}{\textit{Phil. Trans. R. Soc. Lond.}}          
\newcommand{\ptrsla}{\textit{Phil. Trans. R. Soc. London Ser. A}}  
\newcommand{\pw}{\textit{Physics World}}                       %
\newcommand{\rmp}{\textit{Rev. Mod. Phys.}}                     
\newcommand{\rpp}{\textit{Rep. Prog. Phys.}}                    
\newcommand{\ssr}{\textit{Space Sci. Rev.}}                     
\newcommand{\WS}{\textit{World Scientific}}                    %

\providecommand{\newblock}{}

\end{document}